\documentclass[11pt,a4]{report}  
\usepackage{epsf}
\usepackage{graphics}
\usepackage{epsfig}

\newcommand{\beq}{\begin{equation}}
\newcommand{\be}{\begin{equation}}
\newcommand{\eeq}{\end{equation}}
\newcommand{\ee}{\end{equation}}

\newcommand{\bea}{\begin{eqnarray}}
\newcommand{\eea}{\end{eqnarray}}
\newcommand{\bean}{\begin{eqnarray*}}
\newcommand{\eean}{\end{eqnarray*}}

\newcommand{\JGR}{{\em Journal of Geophysical Res.} }

\def\IZ{{\hbox{{\rm Z}\kern-.4em\hbox{\rm Z}}}} 
\def\IR{{\hbox{{\rm I}\kern-.2em\hbox{\rm R}}}}

\def\bei{\begin{itemize}}
\def\eei{\end{itemize}}

\begin{document}
\newcommand{\team}{{\sc IEEC\ }}
\newcommand{\proyecto}{{\sc PIPAER: \\[0.1cm] PARIS Interferometric   Processor Analysis and Experimental Results\ }}

%%%%%%%%%%%%%%%%%%%%%%%%%%%%%%%%%%%%%%%%%%%%%%%%%%%%%%%%%%%%%%%%%%%%%%%%%%%%

%%%%%%%%%%%%%%%%%%%%%%%%%%%%%%%%%%%%%%%%%%%%%%%%%%%%%%%%%%%%%%%%%%%%%%%%%%%%
%%%%%%%%%%%%%%%%%%%%%%%%%%%%%%%%%%%%%%%%%%%%%%%%%%%%%%%%%%%%%%%%%%%%%%%%%%%%
%%%%%%%%%%%%%%%%%%%     END HEADER PART    %%%%%%%%%%%%%%%%%%%%%%%%%%%%%%%%%
%%%%%%%%%%%%%%%%%%%%%%%%%%%%%%%%%%%%%%%%%%%%%%%%%%%%%%%%%%%%%%%%%%%%%%%%%%%%
%%%%%%%%%%%%%%%%%%%     END HEADER PART    %%%%%%%%%%%%%%%%%%%%%%%%%%%%%%%%%
%%%%%%%%%%%%%%%%%%%%%%%%%%%%%%%%%%%%%%%%%%%%%%%%%%%%%%%%%%%%%%%%%%%%%%%%%%%%

%%%%%%%%%%%%%%%%%%%%           TITLE PAGE                    %%%%%%%%%%%%%%%
\vspace{4cm} 
\thispagestyle{empty} 

\ 
 \vspace{4cm} 

\begin{center}  
{\Huge \sc PIPAER} \huge  \\   PIPAER-IEEC-TN-1100/2100 \\  \Large ESTEC Contract No. 14071/99/NL/MM\\ \Large  PARIS Interferometric Processor Analysis and Experimental Results\\ \huge Theoretical Feasibility Analysis\\[3cm] \Large 
{ G. Ruffini, F. Soulat \\ IEEC-CSIC Research Unit \\  Gran Capit\`a, 2-4, 08034 Barcelona, Spain}\\
August 31, 2000
\end{center}

\vfill

%{\small %\copyright\  IEEC-ESA, 2000.    \large This document may only be reproduced in whole or in part, or stored in a retrieval system, or transmitted in any form, or by any electronic, mechanical, or photocopying means with permission of ESA. Furthermore, credits should then be given to the source. }

\clearpage
\thispagestyle{empty}
%%%%%%%%%%%%%%%%%%%%%%%%%   TABLE OF CONTENTS PAGE    %%%%%%%%%%%%%
\tableofcontents \thispagestyle{empty} \clearpage
%%%%%%%%%%%%%%%%%%%%%%%% %%%%%%%%%%%%%%%%%%%%%%%%%%%%%

\chapter{Abstract}

Abstract 

Several  experimental results show that it is possible to extract useful phase information from reflected GPS signals over the oceans. In this work we begin 
the development of  the theoretical background to account for these  results and  fully understand the phenomena involved. This information will then be
used to define and carry out new experiments to evaluate the feasibility of using the phase from reflected GPS signals for altimetric purposes and the
advantages of using interferometric combinations of the signals at different frequencies---the PIP concept. 

We focus on the coherence properties of the signals, including the PIP interferometric combination of phases in the different frequencies.  In this work we
will concentrate on a static, 8 m high receiver  (at least in regards to the simulations), and an infinitely removed static source.  As the ocean moves, the
received field will pick up a random phase.  We want to understand the behavior of this phase, as the goal is to carry out altimetric measurements using phase
ranging.  We will also keep in mind that this random phase carries geophysical information (intuitively, the bigger the significant wave height, the larger the
phase excursions). 

Our simulations are based on the Fresnel integral and use simulated  Gaussian oceans using the Elfouhaily et al. spectrum. The simulation tool, FRESNEL, is
capable of generating time series of the reflected field at different frequencies, and then analyzing their properties.  This software is written in IDL. 

The most important point we need to answer is whether the signal can be tracked in the aforementioned situations. As we show, the PIP combination of the
signals helps clean the signal from noise---the more correlated the signals at different frequencies the more effective is the PIP mechanism. 
The following questions are specifically addressed: 
  \begin{itemize}
\item What is the reflected field spectrum? This is determined by the orbital ocean motion, for a static receiver. We also discuss the moving case and
    show that it depends on the system gain. 
      \item Does the winding number accumulate, or does it average to zero? Winding number refers to phase accumulation. Large phase excursions (many
    cycles)  are seen in our simulations. Our simulations show, however,  that there isn't a preferred direction for phase winding. This means that
    altimetric phase measurements can be accurate. 
\item What are the theoretical values for the average values of the interferometric fields? How do the fields correlate across frequencies? We show that
    the most important factor is the relation between significant wave height and the (real or synthetic) electromagnetic wavelength. 
     \item    What is the coherence time of the signals (the reflected phase)? We show it is longest  for the PIP combination. For rough seas, the correlation
    between the fields (and therefore interferometric coherence) in different frequencies disappears, and the coherence time goes to zero even for the
    interferometric combinations.  In calmer ocean conditions, however, our results indicate that the interferometric combination remains coherent
    while the individual signals lose coherence rapidly. 
      \item What is the structure function of the reflected phase? We see a good fit with a random walk model, with a drift rate proportional to wind speed. 
      \item In what ways is the PIP interferometric signal different and, presumably, superior to the original ones? Coherence, a basic element for altimetric
    purposes. 
     \end{itemize}

Finally, we discuss the robustness of altimetric phase measurements after low-pass filtering the PIP combination. We show that  low-pass filtered PIP data
can provide more robust and accurate altimetric measurements to detect slow-varying geophysical signals in the ocean.
\chapter{Introduction}
This is the sum of Technical Notes PIPAER-IEEC-TN-1100 and PIPAER-IEEC-TN-2100, Theoretical Feasibility Analysis. The {\bf inputs} to this WP are
\begin{enumerate}
\item GMV/IEEC Proposal GMVSA1123/99 (GMV)
\item Relevant literature
\end{enumerate}
The {\bf outputs}  are
\begin{enumerate}
\item This technical note
\item Recommendations for experiments and post-processing procedures
\end{enumerate}
And the {\bf Tasks}:
\begin{enumerate}
\item Review the relevant documentation
\item Establish the worst and best case scenarios for the concept applicability
\item Issue recommendations for experiments and post-processing procedures
\item Prepare this technical note
\end{enumerate}

In addition we analyze some issues relating to the architecture of the proposed PIP instrument.

\chapter{State-of-the-Art, Review}
%d

\section{Historical overview}
%%%%%%%%%%%%%%%%%%%%%%%%%%%%%%%%%%%%%%%
The PIP idea can be traced back to the PARIS concept \cite{Neira_Paris}, where the use of GNSS signals as sources of opportunity for bistatic altimetry was first proposed. The goal of the present  research is to assess the possibilities of   using   the {\em phase}  information of the reflected signal from a PARIS GNSS system for altimetric purposes. In particular, we will study the merits of using the phase from a dual frequency GNSS system in an interferometric fashion.  Other uses of reflected phase data can also be envisioned (surely the temporal behavior of phase fluctuations and other characteristics   contain geophysical information about the sea surface) but we will not explicitly investigate this possibility in what follows.  Rather, we  will, for now, content ourselves with providing   the  background for  discussion of these ideas in future work. This is a natural byproduct of our interest in phase altimetry, as we will see. 

That the GPS signal reflected from the ocean surface can retain coherence under  some conditions has been demonstrated in the past. In \cite{Auber94} we have the first report of a GPS receiver ``locking'' onto a sea-reflected signal.  Motivated by  this somewhat surprising event, tracking of reflected 
signals was later seeked and achieved  for extended periods of time  in a series of experiments with  flights (over the Chesapeake Bay and the Eastern Shore of Virginia) at 
an altitude of up to 5500 m \cite{Garrisonb}.  This was accomplished using an Ashtech Z-12 off-the-shelf receiver with a LHCP nadir looking 
antenna, and it was  observed that {\em carrier} lock was obtained. This particular receiver  would not have  been able to track 
otherwise.

On the other hand, several groups have to date successfully analyzed reflected data in order to correlate the direct and reflected signals with clean code 
replicas. In the experiment carried out by ESA in September 1997  two receivers  were employed,   operating  independently of each other, one tracking the direct signal  with an 
up-looking antenna, the other recovering  the  reflected signal via  a down-looking LHCP 9 dB helical antenna over an 18 m bridge near Rotterdam (The Netherlands), the ``Zeeland 
Brug". After amplification, each of the signals was sent to a different GPS receiver (GEC Plessey builder kit 2).  The receivers  down-converted the signals to IF (4.309 MHz, with a bandwidth of 1.9 MHz). The IF signals were sent directly to a high-performance 
sampling card, which sampled the data at 6.25 MHz  with  2 MHz bandwidth. They were then cross-correlated with replicas of the signal adjusted for 
Doppler\index{Doppler effects} due to satellite motion and with 20 ms duration (20 C/A code periods).  Using the correlation delay data from the direct 
and reflected signal, this group  solved for the height of the bridge over mean surface (which depended on the tide) as well as a hardware bias constant via a  least 
squares fitting procedure.  In addition, the receivers recorded standard RINEX files, 
including SNR. The GPS signals were deemed usable if the SNR of the reflected signal was greater than 6 to 9 dB, and if the geometry was 
favorable (visibility and multi-path were rather complicated in some geometrical configurations, which were then excluded). Significant Wave Height  during the experiment was 
around 1.4 meters (equivalent to a height standard deviation $\sigma_\zeta$ of about 35 cm), and the altimetric performance was rather poor, as only the  C/A code  could be used for correlation (in general, a 1\% of the chip length is 
assumed to be the optimal precision for one measurement---this is 3 m for C/A code). A very good 
description and  further details can be found in \cite{Caparrini98}.

The IEEC Earth Sciences Department has been able to reproduce the analysis of the Bridge Experiment data  with similar results. The analysis tools for 
the generation of the GPS signals and their correlation with the received ones  have been developed in Matlab.

Another aspect of the Bridge Experiment analysis carried out by ESA was the attempt to extract phase information from the data.   The first attempt of 
phase processing  was an evaluation of Doppler frequency based on zero-crossings counting \cite{Caparrini98}.  The approach was to first multiply the 
signal  by  an appropriate clear replica of the PRN code of a satellite in view. It was then expected that a peak in the Fourier transform of the resulting  
signal would  appear at the IF plus the Doppler frequency.  The direct approach was not deemed feasible, however,  due to the limited length of the signals (only 
10 ms long),  which did not allow for enough spectral resolution. The signal multiplied by the replica was instead  first filtered with a rather narrow filter in 
order to eliminate all but the expected Doppler-shifted component.  After that, frequency was estimated using the zero-crossings count for each 10 ms stream of data, with 
rather good results for both the direct and the reflected signal. This suggested that the reflected signal still contained a certain degree of coherence.  A 
second study was then devoted to the measurement of the relative tide height (that is, the difference in the height measurements at different times through 
the difference in the number of zero-crossings between the direct and the reflected signals).   The reasoning  was that if the phase could be measured at all, it 
would be only up to a phase bias. If the bias is assumed to be constant over some short measurement times, then the change in the phase between those 
times would be related directly to the change in delay between the direct and the reflected signals.  This calculation, however, could not be carried through 
because the data was too noisy and the intervals between data measurements too far apart in time.

\section{The goal of this research}
As we have seen,  several  experimental results show that it is possible to extract useful phase information from reflected signals. In the analysis sketched below we will begin  to develop the theoretical background to account for these  results. The goal of this work is to continue this development and analysis of past work to fully understand the phenomena involved. This information will then be used to define and carry out new experiments to evaluate the feasibility of using the phase from reflected GPS signals for altimetric purposes.

We will focus on the coherence of the signals, including the interferometric combination of phases in the different frequencies.  In this work we will concentrate on a static, 8 m high receiver  (at least in regards to the simulations), and an infinitely removed static source.  As the ocean moves, the received field will pick up a random phase.  We want to understand the behavior of this phase, as the goal is to carry out altimetric measurements using phase ranging.  We will also keep in mind that this random phase carries geophysical information (intuitively, the bigger the significant wave height, the larger the phase excursions). 

The PIP concept concept consists in combining two phases from the signals at different frequencies. Imagine a very smooth and slowly changing ocean. By combining the two phases we obtain a slowly changing interferometric phase. In some sense we would expect that this interferometric phase will  be less sensitive to small ripples in the ocean and thus provide a  more robust ranging tool for altimetry.   

Several implementations of this concept can be envisioned: a static receiver on a coast or bridge to monitor tides or floods as well as  sea state. Or  deployment on boats, aircraft and spacecraft for global altimetric or oceanographic measurements. 
Carrier altimetric signals over the oceans, even when seen by a fast moving Low Earth Orbiter, will  mostly be low frequency. A strong topographic  signal over the oceans would be a slope of 1 meter every 100 km. As seen by the LEO (traveling at 7 km/s, say), this translates into 0.35 Hz in L1, a rise of 5 cycles in 100/7 seconds. In L25, this becomes 0.01 Hz.

On the other hand, the roughness and motion of the sea will induce high frequency ``jitter'' on the received signal. This is a nuisance for altimetry, and this is where we expect the PIP concept to bring added value, as the interferometric combination should be less sensitive to these effects. It is appropriate to recall here the old ado, ``What is noise to some is signal to others.'' This jitter may contain very useful information.

In the author's mind, the most important point we need to answer is whether the signal can be tracked in the aforementioned situations. As we will see, the PIP concept will help clean the signal from noise. For the most part,  we will focus  in this work  on the low altitude, static receiver situation. The moving cases at higher altitudes will be left for future work, as we will discuss in the last Chapter. 
%Esta claro que tendremos que buscar senyales en phase de tipo geophysico.
%Bertrand hablaba de que una senyal fuerte seria de un metro en 100 km, o una
%senyal de 5 ciclos en 100/7 segundos ( una velocidad LEO de 7 km/s), o sea de
%0.35 Hz. El problema es distinguir esta senyal del ruido causado por la
%rugosidad del mar. Estos 5 ciclos en 100/7 segundos  se convierten en 0.2
%ciclos
%con la frequencia L25.  O sea, la senyal geofisica el L1 estaria a 0.35 Hz,
%pero
%en la frecuencia interferometrica esta a solo a 0.01 Hz, que es mucho mas
%filtrable.

%Creo que la magia del PIP consistira en que las cosas a pequenya escala del
%mar
%estan decorreladas en las dos frequencias, mientras que las grandes correlan
%mejor. Y que por eso, filtrar al final te da lo que quieres en el caso
%interferometrico.

In this report we will address, among others, the following questions:
\begin{enumerate}
\item What is the reflected field spectrum?
\item Does the winding number (to be described below) accumulate, or does it average to zero?
\item What are the theoretical values for the average values of the interferometric fields? That is, how do the fields correlate across frequencies?
\item What is the coherence time of the signals (the reflected phase)?
\item What is the structure function of the reflected phase?
\item In what ways is the interferometric signal different and, presumably, superior to the original ones?
\end{enumerate}

 \chapter{Background and Applicable Models}

The Rayleigh criterion\index{Rayleigh criterion} defines  a surface to be  smooth if $\sigma_\zeta< \lambda/(8 \cos \theta)$, where $\sigma_\zeta$ is the surface height standard deviation,  $\theta$ is the incidence angle with respect to the normal to the  surface. A more stringent  condition is provided by the Fraunhofer criterion\index{Fraunhofer criterion}, which is used to define the far-field distance of an antenna, 
$\sigma_\zeta< \lambda /(32 \cos \theta)$. A surface becomes smooth under two conditions: $\sigma_\zeta\sim0$ or $\theta\sim 90^o$. The  effective roughness of the surface is therefore $\sigma_\zeta \cos\theta$  \cite{Beckmann}.

Several aspects must be considered to understand the scattering  phenomenon.  We will examine two limiting  situation models. In one case, we can assume that the surface is made up of many independent   mirrors (specular points) whose contributions to the field   add up incoherently. The other limit is that of a surface which deviates from flatness only slightly. These two cases must be treated differently.  The goal in both cases will be to understand the behavior of the received phase and modulus. Let us keep in mind  in the back of our heads that, at the end, we should look at the {\em relative phase} between the two available wavelengths, which should be a statistically better behaved quantity than either of the two phases. 
\section{Statistics of the field}
In this subsection we  briefly review some of the results in \cite{Beckmann}, concerning the sum
\beq
U=r e^{i\psi} =\sum_{j=1}^n   e^{i\phi_j}.
\eeq
This is relevant to understand what happens when we sum the fields in a  rough reflection  situation. We state the result for two illuminating conditions:
\begin{enumerate}
\item A uniformly distributed phase from $-\pi$ to $\pi$. 
\item A normal phase distribution with standard deviation $\sigma$.
\end{enumerate}
In the first case the resulting phase has again a uniformly distributed phase, but the modulus has the so-called Rayleigh  distribution $p(r)={2r\over n} \exp (-r^2/n)$---and therefore has a non-zero average (the average is proportional to $n^{1/2}$). In the second case we find that the components of the resulting phasor are normally distributed (which is a quite general result following directly from the Central Limit Theorem) and that 
\beq
\langle r^2 \rangle = n^2 e^{- \sigma^2} +n(1-e^{- \sigma^2}).
\eeq
In this beautiful expression we can see    the coherent and incoherent contributions to the field modulus. Beckmann summarizes:
\begin{quote}
Outside a narrow cone (or wedge) about the direction of specular reflection, the amplitude of the field scattered by a rough surface is always Rayleigh-distributed; if the surface is very rough, and grazing incidence is excluded, the amplitude of the scattered field is Rayleigh-distributed everywhere. 
\end{quote}
The specularly scattered field is  composed of a coherent component and a random, Hoyt-distributed component. When the surface is very rough, the latter becomes incoherent and the former vanishes.  In fact, if the surface height distribution is normal with deviation $\sigma_\zeta$,  then 
\beq
\langle r^2 \rangle =n^2 e^{-( 4\pi \sigma_\zeta\cos\theta /\lambda )^2} + n(1- e^{-(4\pi \sigma_\zeta\cos\theta /\lambda)^2} ).
\eeq
Thus, we see that the magnitude of the reflected field should depend mainly on the ratio of significant wave height to wavelength.  We will obtain a related result through more sophisticated analysis below.

Some verification of these ideas via simulations  can be found in \cite{Daout99}, where the authors, using a 1-D bistatic scattering simulator, reproduce the mentioned aspects of the phase statistical distribution.
The surface is there simulated by facets, and the Kirchhoff theory for the scattering field from each facet is used to accumulate the rays with regard to their phase.  The phase histograms they show verify very clearly the ideas just discussed. In particular, that the signal is very coherent at low elevations.

\section{The Fresnel-Huygens-Kirchhoff integral for the field}
This is basically the scalar Kirchhoff approximation, valid for surfaces with large radii of curvature compared to wavelength.
Let the incoming field be described by\footnote{Here we do a scalar treatment, think of $U$ as a component of the field.} 
\beq
U_{o}(p)= {e^{ikr} \over r}.
\eeq
The Fresnel integral for the scattered field is (see \cite[p. 380, eq. 17]{BornWolf})
\beq
U(p)= {-i \over 4 \pi} \int {\cal R}\cdot { e^{ik(r+s)} \over rs} \, (\vec{q} \cdot \hat{n})\,  dS .
\eeq
 The vector $\vec{q}=(\vec{q}_\bot, q_z)$ is  the {\em scattering vector}\index{scattering vector}: the vector   normal to the plane that would specularly reflect the rays in the direction we are looking. This vector is a function of the incoming and outgoing unit vectors $\vec{n}_i$ and $\vec{n}_s$, $\vec{q}= k(\hat{n}_i - \hat{n}_s)$.  The scattering vector is related to the specular angle $\beta$ through $\cos \beta=\hat{z}\cdot \hat{q}/q$. 
Note, as an aside, that in the nadir case all wind direction dependence  disappears (surface anisotropy). Changing wind direction is akin to performing a rotation in the surface, and the integral is invariant under rotations in the nadir case. The nadir case is special because $R[\vec{q}(\vec{x})]=\vec{q} (R[\vec{x}])$.

 \section{Geometrical Optics}
This is the process that dominates in specular scattering when the frequency is large (i.e., the wavelength is small compared to the wave height and to the surface correlation length).  The case of L-band scattering from the ocean is at best borderline ``high-frequency'', and we are carrying out some separate studies within the framework  of GNSS-OPPSCAT\footnote{ESA Contract 13461/99/NL/GD,  Utilization of Scatterometry Using Sources of Opportunity.} to understand the relevance of this approximation in such circumstances. To this end, we have developed the {\sc speckles} software tool, written in IDL (Interactive Data Language). In Geometrical Optics   the surface height standard deviation is assumed to be at least of the order of the electromagnetic wavelength.  According to the analysis in \cite{Beckmann}, the resulting wave will be largely incoherent. There is an important ingredient  in our situation, however, and that is because we have a Woodward Ambiguity Function (WAF) at our disposal to select a given surface patch, as we will see shortly. This means that we can filter field contributions from a rather small area. 

In the Geometrical Optics approximation to the Kirchhoff theory for electromagnetic 
scattering, the physical picture can be understood in terms of a specular point model.  That is, the field at the receiver is the superposition of the fields generated by a 
number of "mirrors" (not flat mirrors, thought, rather parabolic caps) on the scattering surface which are oriented in the correct manner.  %The size of each mirror is of the order of the first Fresnel 
%zone (the Fresnel zone is defined as the set of points near the specular with ranges to the receiver equal or less that of the specular range plus a wavelength).  The size of this area depends on the bistatic geometry.  
The radiation from each specular point is as coherent as the incoming radiation, but there is no coherence in the phase relationship between the radiation out-coming from the different mirrors if the surface height distribution has a large range.   The result 
is that there is power at the receiver, but that the phase of the received signal is randomly distributed and the power is simply proportional to the number of scatterers.  The behavior of the resulting phase will depend on
the geometry, on the number and distribution of mirrors, and also on the temporal variation of these quantities.   The goal of this section  is to understand this relationship and quantify it. 

It is useful here to stop and give a physical picture of the specular point model---one of the two ingredients in Geometrical Optics. We will treat here the radiation coming from each mirror separately, as was mentioned above.  Since the size of each mirror is small compared to the Fresnel zone, we can use the Fraunhofer approximation (plane waves all the way through) to estimate the out-coming field. 

\begin{figure}[h!]
\hspace{2.5 cm}
\epsfxsize=80mm
\epsfysize=80mm
\epsffile{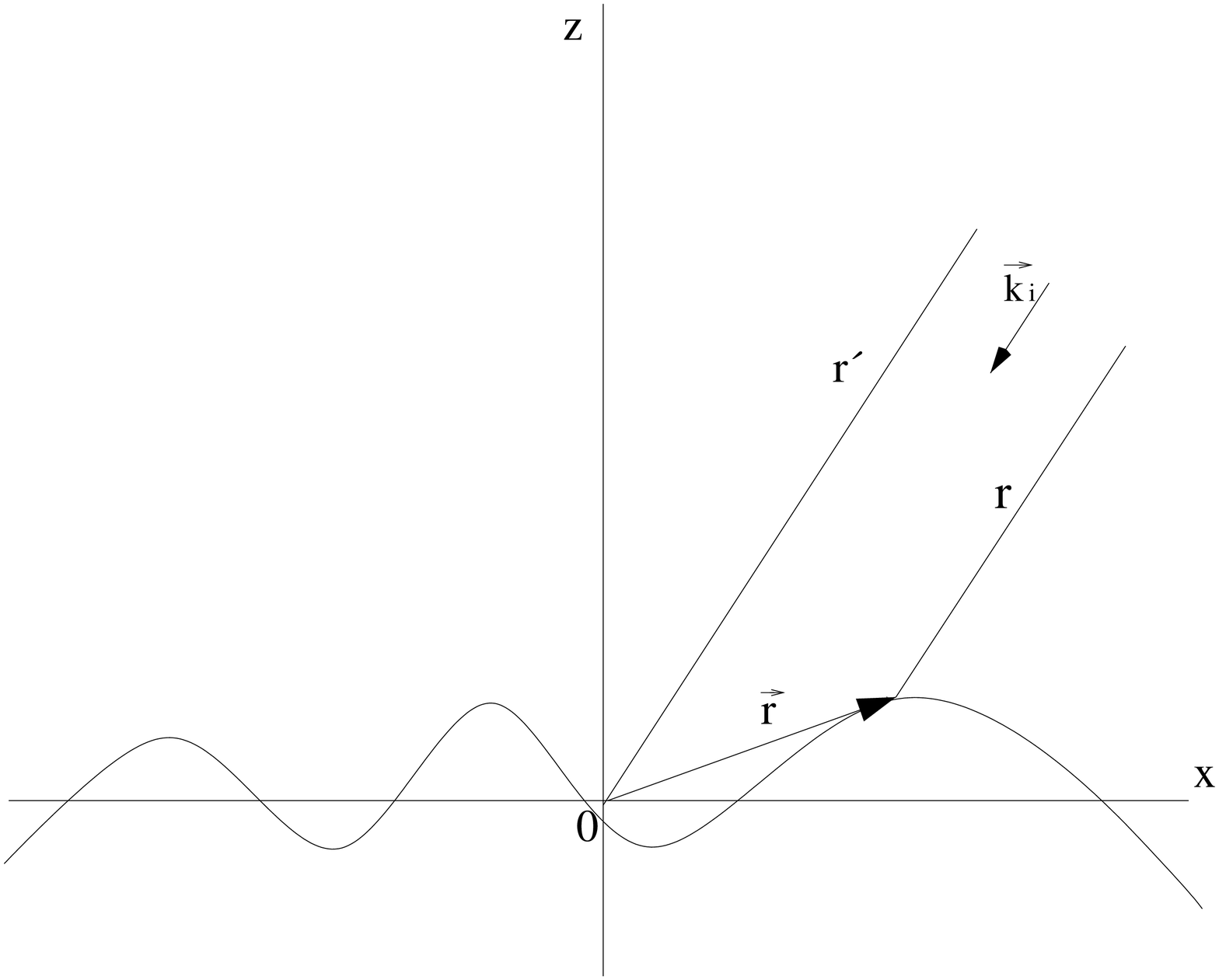}
%\resizebox*{9cm}{9cm}{\includegraphics{fresnel_V13.eps}}
\caption{ \label{vectors} Vector geometry for scattering. }
\end{figure}

Recall that the Fresnel integral for the scattered field is 
\beq
U(p)= {-i \over 4 \pi} \int {\cal R}\cdot { e^{ik(r+s)} \over rs} \, (\vec{q} \cdot \hat{n})\,  dS .
\eeq
In the far field (both emitter and receiver are very far compared to the size of the scatterer, i.e, the size of the scatterer is much smaller that the Fresnel zone), and since our integral is near a specular point with $\hat{n} \approx \hat{q}$, we find  that 
%$ dS= {d^2 \vec{\rho} \over  \vec{n}\cdot \hat{z}} = d^2\vec{\rho}{  q \over |q_z|}$)
\beq
U(p)={-i    \over  4\pi}   {e^{i k (r'+s')} \over r's'}\int {\cal R}\cdot e^{-i \vec{q}\cdot \vec{r}} \, (\vec{q} \cdot \hat{n})\, dS,
\eeq
see Figure \ref{vectors} for a pictorial definition of the involved vectors. 
Here all we have done is to approximate $ k(r+s)\approx k (r'+s')-\vec{q}\cdot \vec{r}$.  Now, let us focus on a single specular point: imagine that there is only one specular point on the surface.  This means $\vec{q} \cdot \hat{n}=q$. To perform the integral let us use a coordinate system in which the $z$ axis is parallel to $\vec{q}$. In this coordinate system the tangent plane at the specular plane is therefore parallel to the $x$-$y$ plane. Hence,  $dS= {d^2 {x} \over  \vec{n}\cdot \hat{z}}=d^2 {x}$. We need to compute
\beq
I=q \int {\cal R}\cdot e^{-i q \,  z(x,y)} \, d^2 {x} , 
\eeq
where 
\beq U={-i    \over  4\pi}   {e^{i k (r'+s')} \over r's'} I. \eeq 
Now we use the stationary phase approximation. The idea is that as $q$ gets larger the integral gets contributions only very near the specular point---the contributions farther out cancel out. Then,
\bea
I    &\approx & I_0=q \, {\cal R}_{spec} e^{-i q \,  z_{spec}} \int e^{-i {q\over 2} \, \left. \nabla_{ij}^2 z \right|_{spec}\, x_ix_j}  \, d^2 {x}\\
         &=&  q \, {\cal R}_{spec} e^{-i \vec{q}\cdot \vec{r}_{spec}}  {2 \pi \over -i {q} \,   \det^{1/2} \left(  \left. \nabla_{ij}^2 z \right|_{spec} \right)} \\
         &=&   i \, {\cal R}_{spec} e^{-i \vec{q}\cdot \vec{r}_{spec}}   {2 \pi \over   \det^{1/2} \left( \left. \nabla_{ij}^2 z \right|_{spec} \right) }.
\eea 
This result, that the square root of an inverse determinant follows from a Gaussian integral is a much used  result  in Quantum Field Theory (see for instance, Quantum Field Theory by L. H Ryder, Cambridge Univ Press 1996  for a derivation). It can be seen to follow from extending
$$
\int e^{-\alpha x^2} \, dx =\sqrt{\pi\over \alpha}
$$
to higher dimensions.
Now, it is not hard to show that  $\det^{1/2} \left( \left. \nabla_{ij}^2 z \right|_{spec} \right)^{-1}= \sqrt{r_1r_2}$---the determinant just yields the products of the radii of curvature at the specular point: the determinant is invariant under coordinate transformations of the surface, so we can use a coordinate system along the principal directions---the surface is then described as a simple parabola. The result is then immediate (the radius of curvature is the inverse of the second derivative in such coordinates and  the matrix becomes diagonal in such a coordinate system, so the computation of the determinant just yields the product of the diagonal terms).  Finally,
\bea
U_0(p) &=& {-i    \over  4\pi}   {e^{i k (r'+s')} \over r's'} \cdot 
i \, {\cal R}_{spec} e^{-i \vec{q}\cdot \vec{r}}   {2 \pi \over   \det^{1/2} \left( \left. \nabla_{ij}^2 z \right|_{spec} \right) }\\ 
&=&  {{\cal R}_{spec}\over 2} {e^{-i k (r_{spec}+s_{spec})} \over r_{spec} s_{spec}} \sqrt{r_1r_2} \label{eq310}
\eea
The  coefficient ${\cal R}_{spec}$ depends on the Fresnel coefficients and on the local geometry---we will discuss this below. 
The cross section is given by the ratio of the resulting field squared  divided by the field squared  at the surface (given above) times $4\pi r^2$. The result is 
\beq
\sigma= {4\pi r^2 |U|^2 \over |U_o|^2} = |{\cal R}_{spec} |^2\pi r_1r_2 .
\eeq
For all this to work, some requirements have to be met on top of those from the Kirchhoff approximation. Basically, this is a high frequency approximation---$q$ has to be large. Let us look at things in 1D, along one of the radii of curvature. The integral is essentially of the form
\bea
J &=&\int_{-l_x}^{l_x} dx\, e^{iq x^2/r_x} \\
  &=& \sqrt{r_x/q}\int^{l_x\sqrt{q/r_x}}_{-l_x\sqrt{q/r_x} } e^{iu^2} du .
\eea
For the integral to become $\sqrt{\pi}$, we need $l_x\sqrt{q/r_x}$ to be large---of the order of 10.
% Let us rewrite $l_x=\cos \nu \cdot r_x$.  Then, we need
%\beq
%\cos\nu \cdot \sqrt{qr_x} >10,
%\eeq
%say.
% As  the specular point size is around the radius of the tangent sphere, the approximation is ok.
   The expression for the incoherent sum of all the specular points is now immediate from Equation \ref{eq310}, given a large height deviation. This expression for the field is also useful for a not-so-incoherent sum of the fields---we will return to this point below.
%%%%%%%%%%%%%%%%%%%%%%%%%%%%%%%%%%%%%%%%%%%%%%%%%%%%%%%%%%%%%%%%%%%%%%%%%%%%%%%%%%%%%%%%
\subsection{Corrections to the Geometrical Optics approximation: frequency dependence}
In this section we add a comment on the next order corrections to the Geometrical Optics approximation. This is not a crucial section for the work at hand and can be taken as a small aside, but we believe it will be useful in future work. 

In reality the surface is not a parabola and other terms appear. These are frequency dependent corrections to the field. In order to obtain the corrections we need to expand the integrand as a Taylor series beyond the well-behaved parabolic term. The basic idea for this type of computation is that 
$$
I(b)=\int e^{-ax^2+bx} dx = e^{-b^2/4a} \sqrt{\pi/a},
$$
and that 
$$
\left. {d^n \over db^n} I(b) \right|_{b=0}= \int e^{-ax^2+bx} x^n dx.
 $$
From this we can compute
$$
\int e^{-ax^2} x^4 dx = {12\over (4a)^2}  \sqrt{\pi/a}.
$$
The expansion is 
\beq
I=q \int {\cal R}\cdot e^{-i q \,  z(x,y)} \, d^2 \vec{x}= I_0+I_4+...,. 
\eeq
where
\beq
I_4=q \, {\cal R}_{spec} e^{-i q \,  z_{spec}} {-iq \over 4!} \left. \nabla_{ijkl}^4z\right|_{spec} \int \, x^i x^jx^kx^l  e^{-i {q\over 2} \, \left. \nabla_{ij}^2 z \right|_{spec}\, x_ix_j}  \, d^2 \vec{x}.
\eeq	
%We now want ot compute the next relevant correction (odd powers of $x$ vanish),
Now, 
\bea
I_4 &=& q \, {\cal R}_{spec} e^{-i q \,  z_{spec}} {-iq \over 4!} \left. \nabla_{ijkl}^4z\right|_{spec} \int \, x^i x^jx^kx^l  e^{-i {q\over 2} \, \left. \nabla_{ij}^2 z \right|_{spec}\, x_ix_j}  \, d^2 \vec{x}\\
&=& q \, {\cal R}_{spec} e^{-i q \,  z_{spec}}
{-i q \over 4!} \left. \nabla_{ijkl}^4z\right|_{spec} + \nonumber \\ 
&& \;  \left( \left. -iq \nabla_{ij}^2 z  \right|_{spec}\right)^{-1} 
\left(  \left. -i q \nabla_{kl}^2 z \right|_{spec}\right)^{-1}   {2 \pi \over  -i q  \det^{1/2} \left( \left. \nabla_{ij}^2 z \right|_{spec} \right) }
\eea 
and 
\beq
I_4 =   {\cal R}_{spec} e^{-i q \,  z_{spec}} {-2\pi \sqrt{r_1r_2}\over q\,  4!} {\cal R}_{spec} e^{-i q \,  z_{spec}}   
 \left. \nabla_{ijkl}^4z\right|_{spec}  \left( \left.  \nabla_{ij}^2 z  \right|_{spec}\right)^{-1} 
\left(  \left.  \nabla_{kl}^2 z \right|_{spec}\right)^{-1}    
\eeq
In a future implementation of {\sc speckles} (see discussion below), we would like to include these second order effects.  These effects can also be accounted for theoretically using, e.g., Gaussian statistics. At the end the final expression must depend only on the usual parameters. Note that the $I_4$  has units of  length. Thus, aside from constants, these higher order corrections may be of the form $\sigma_\zeta \over l \cdot  q$, say.  In this last expression we see that this correction is indeed inversely proportional to frequency, as it should, since Geometric Optics is a high frequency limit.
\section{The WAF zone}
If the signal is filtered by 
cross-correlation with several C/A (or P) code periods at the delay and frequencies corresponding to the specular zone, for instance, only the surface patch  in the first  Delay and Doppler zone will contribute (let us call it the ``WAF zone'' to associate as well as distinguish this concept from the concept of  Fresnel zone). This is described by the equation
$$
\mbox{SNR}(\tau,f_c)=  { 1\over kT_S B_D} \cdot {\lambda^2\over (4\pi)^3} \int {P_t G_t G_r \over R_1^2 R_2^2}  \sigma^0 _{rt}\, \chi^2(\vec{\rho},t,\delta\tau, \delta f)\, dA. 
 $$
which is a slight refinement of a similar equation in \cite{Zavorotny99} and which can be found in the review \cite{Ruffini99}. The important thing to keep in mind is that the support in the integrand,  the WAF zone, is the intersection of four spatial zones:
\begin{enumerate}
\item The receiver antenna footprint.
\item The annulus zone defined by the $\Lambda^2$ function in $\chi^2$.
\item The Doppler zone defined by the $|S|^2$ function in $\chi^2$\index{Doppler zone}.
\item The scattering cross section coefficient $\sigma_0$.
\end{enumerate}

The WAF, through its support,  selects  a given portion of the surface from which power is measured. 

   To get a feeling for these numbers, the area involved in the reflection is of the order of 10 km$^2$ for an aircraft and 1000 km$^2$ for a Low Earth Orbiter. For $h >> c\tau$, where $\tau$ is the chip length (1 $\mu$s for C/A code) and $h$ the receiver height), the one-WAF (or one-chip here)  area can be approximated by
$A_{WAF}={2\pi h  c\tau / \cos^2 \theta}$.  {\em Note that $A_{WAF}$  increases only linearly with altitude. }

\begin{figure}[b!] 
\hspace{1.5cm} 
\epsfxsize=120mm 
\epsffile{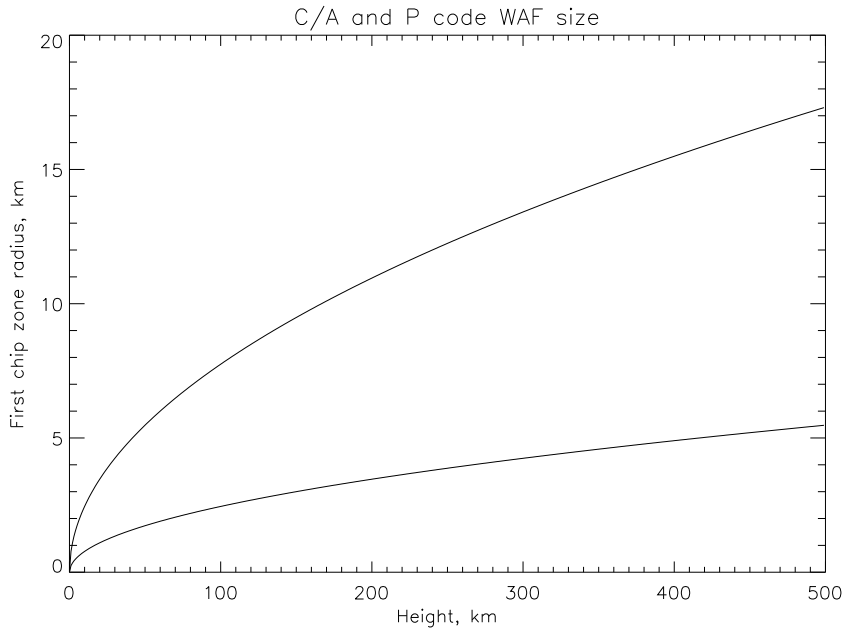} 
\caption{ \label{chipsize} Nadir case first chip radius size for C/A and P codes.  This is for a static situation, no Doppler filtering affects the result (based on the simple formula $s=\sqrt{2*h*\tau_c}$), where $\tau_c$ is the chip length in meters (300 m for C/A code).} 
\end{figure}

We can add another element at this point. In the above discussion, the WAF is assumed to result from choosing a given delay and Doppler in the reflected signal. However, as time passes the signal reflected by a given surface element cannot be characterized by a fixed Delay and Doppler: both will change as the situation evolves. It is clear, however, that a more sophisticated WAF algorithm can be devised that focuses on that specific surface patch. This observation opens the door to longer integration times, which can be useful for altimetric purposes.

\chapter{Simulation Tools: \sc fresnel and speckles}

These are two modeling tools we have developed to understand, via a computer simulation, some of the characteristics of the reflected field. {\sc fresnel} models the reflected field (just the carrier part after removing the source time-dependence) using an ocean model and direct integration of the Fresnel-Huygens-Kirchhoff integral described above.  We have carried out several types of simulations: a slowly vertically moving receiver, a static receiver, a realistic ocean and a so-called chaotic ocean.  We will focus here on the static realistic ocean case.

{\sc speckles} has been developed with two goals in mind. We will eventually attempt to compute the reflected field  using a GO method: finding first the specular points and integrating just around these to sum up the resulting field. We will just outline here some of its characteristics and initial results.  We have found that this is very difficult to do (in general, comparisons with the Fresnel integral are poor). We have seen, however that the number of specular points correlates reasonably well with the reflected Fresnel field, specially in  rough ocean simulated conditions. We are attempting to understand this phenomenon better, and to study its implications: GO, in some form of another,  is presently the leading model used for the analysis of reflected power in GPSR. It  is a simple method, and experimental results to date seem to validate it. We would like to understand its somewhat puzzling high performance, since L-band is not really  high-frequency in the ocean surface case. That is, GO is not, strictly speaking applicable in the whole range of ocean conditions with an L-band instrument---the wavelength in neither large of short enough. Nonetheless, some semi-empirical modifications of this model are currently being successfully used. We would like to understand this success and, more importantly, its limitations.   We will briefly mention some of this work, although it is not central to the study.    

\section{\sc fresnel}
This routine ( written in Interactive Data Language---IDL) integrates the field from the ocean surface below, using nadir incidence. The receiver is hovering a few meters over the water. The ocean model is Gaussian (we use a routine from B. Chapron using the Elfouhaily spectrum) \cite{Elfouhaily97}, ``chaotic'' (random heights, moving) and a non-Gaussian simulations are  planned for future work. The software simulates the reflected complex field at L1, L2 and L5 frequencies.  The goals of this part of the work were to show that the simulation is possible, to compare the behavior of the field in the different cases, and to assess the potential of the PIP idea.

\subsection{Random ocean}
The random ocean is generated by choosing a height standard deviation. The code assigns a random height with the desired characteristic to each 10 cm by 10 cm square of the reflecting surface. In addition, a velocity random field is chosen, and each square moves up and down with  a chosen period. 

\subsection{Elfouhaily spectrum}
The procedure to compute the ocean state, based on the IDL routine kindly supplied by B. Chapron, is, starting from the energy  spectrum, to generate a plausible random spectrum. To do this, the square root of the energy spectrum is taken, and a uniformly random phase appended to it.  The appropriate time dependence is also added to this random phase---using the dispersion relation for gravity ocean waves.   Then, the inverse Fourier transform is taken. Using this procedure, a moving, random ocean with  the appropriate spectral characteristics is generated.  Using a uniform random phase distribution in the different frequencies means that the resulting ocean is Gaussian. This can be changed, of course, but we have not attempted to analyze the subtleties of non-uniform phase distributions in this work.

Generally the sea state is characterized by its power spectrum. As the sea evolves in a random process, the spectral density of sea elevations is obtained by the Fourier transform of the autocorrelation function of the elevations $\zeta(x,y)$. The spectral density can also be estimated by the Fourier transform of one realization of $\zeta(x,y)$ [Blackman and Tukey, 1958]:
$$
F(k_x,k_y) = |TF[\zeta(x,y)]|^2
$$
If we determine the matrix $\phi$ of random phases uniformly distributed between 0 and $2\pi$, the sea height at the point $r = (x,y)$ is:
$$
\zeta(r) = TF^{-1}[\sqrt{F(k_x,k_y)}e^{i\phi}]
$$
Thus the probability density functions of heights and slopes are Gaussian. For this study we use the unified spectrum of \cite{Elfouhaily97}. This is just a consequence of the Central Limit Theorem applied to sums of harmonics.

\subsection{Field calculation}
The field is calculated as a complex number using the Fresnel integral expression:
\beq
U(p)= {-i \over 4 \pi} \int {\cal R}\cdot { e^{ik(r+s)} \over rs} \, (\vec{q} \cdot \hat{n})\,  dS .
\eeq
To be precise, we Fraunhofer-expand the part of the integrand corresponding to $s$ the distance to the infinitely far transmitter. That is, we write $s=ks'-\vec{k}_{in}\cdot \vec{x}$ in the phase part of the integrand, where $\vec{x}$ denotes the position of the scattering point. Thus, the integrand becomes,
\beq
U(p)= {-i \over 4 \pi} {e^{iks'}\over s'}  \int {\cal R}\cdot { e^{ikr-  \vec{k}_{in}\cdot \vec{x}} \over r} \, (\vec{q} \cdot \hat{n})\,  dS .
\eeq
We actually throw out all the $s'$ dependence  (in effect assuming an incoming plane wave, as well as on the Fresnel coefficient ) and obtain
\beq
U(p)= {-i \over 4 \pi}  \int {\cal R}\cdot { e^{ikr-  \vec{k}_{in}\cdot \vec{x}} \over r} \, (\vec{q} \cdot \hat{n})\,  dS .
\eeq
This means that the outgoing field would be a plane wave if the surface were  flat, with unit modulus.

The ocean surface is divided into squares 10 cm wide (or less), and an area of up to 200x200 meters is integrated. For the most part we will concentrate on the first chip zone, with radius $r=\sqrt{2*h*\tau_c}$.    

The resulting field is a complex number in this scalar treatment, $U(p(t))=re^{i\phi(t)}$. In order to evaluate the phase as the receiver moves up we take
\beq
\delta\phi_i = \ln (w(t_i)/w(t_{i-1}))
\eeq 
with 
$w(t)=e^{i\phi(t)}$. Summing the $\delta\phi_i $ yields the overall phase.  This method is related to the definition of {\em winding number} of a curve around the origin in the complex plane:
\beq
I=\oint_\gamma {dz \over z} =\oint_\gamma d(\ln z) = \sum_i \ln {z_i\over z_{i-1}}.
\eeq 
A full cycle, 360 degrees, is equivalent to 19 (L1), 24 (L2), 25 (L5), 86 (L12) or 568 (L25) cm.
We will return to this definition shortly.

\begin{figure}[b!] 
%\hspace{2.5cm} 
\epsfxsize=150mm 
\epsffile{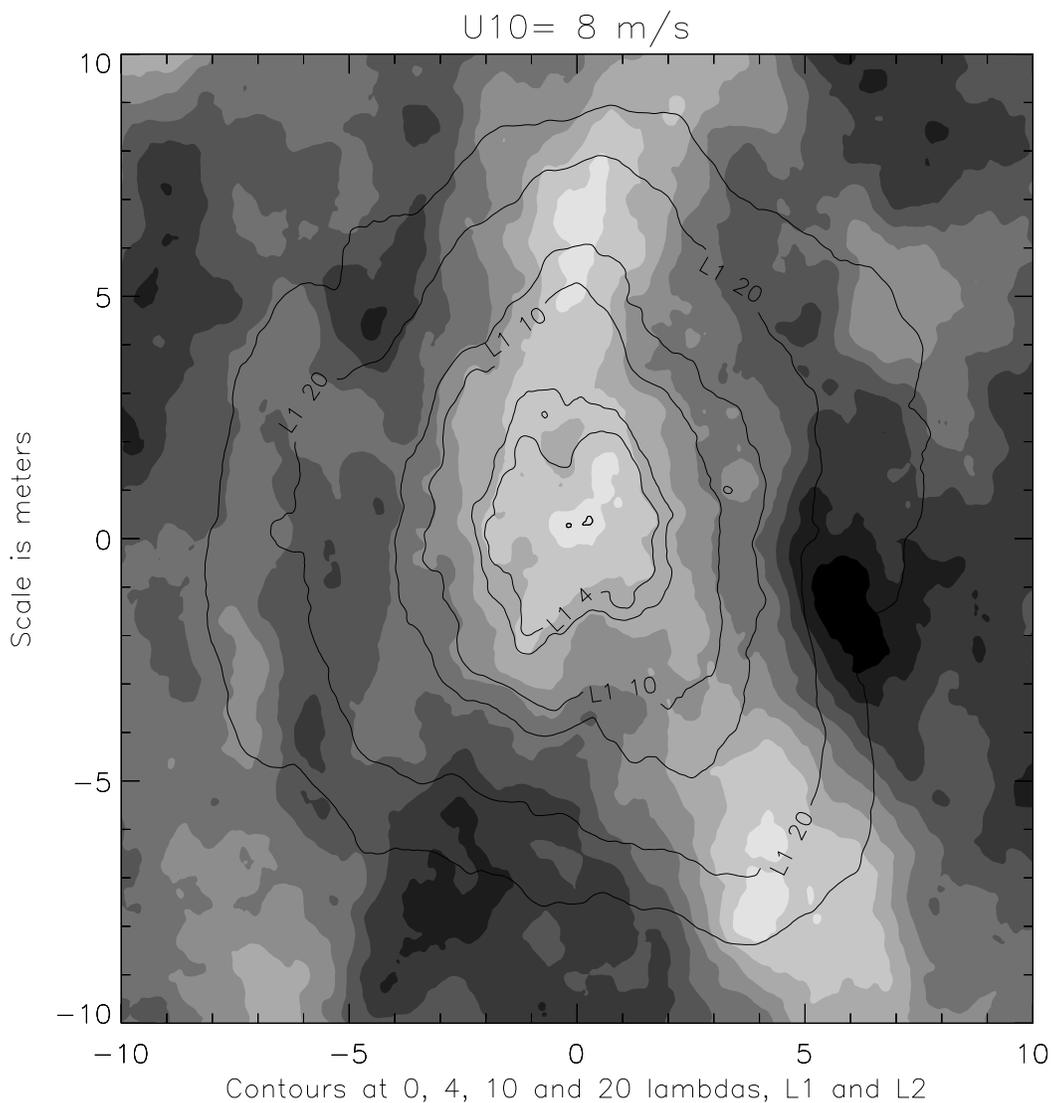} 
\caption{ \label{contours}Fresnel contours for the 10 by 10 surface patch under and 8 meter high receiver. U10=8 m/s. The colored contours represent the ocean instantaneous topography, with higher areas represented  brighter. The line pairs are iso-lambda delay curves for L1 and L2 (4, 10 and 20 lambdas are shown).  } 
\end{figure} 

%Longer runs, using now the area of the first chip (about 90 meters for a 10 meter high receiver, $\sqrt{2*h*\tau_c}$), confirm these results:

%{\small \begin{verbatim}
%iteration:      65       57.702209-- Minutes elapsed
%       40.1432     -43.3273      10.2367max, min, swh, in cm
%       155.18068 L1 average  jitter, degrees       8.2084289---> cm
%       85.239951 L2 average  jitter, degrees       5.7863549---> cm
%       79.093328 L5 average  jitter, degrees       5.6025421---> cm
%       71.106041 L12 average  jitter, degrees       17.036121---> cm
%       8.8042465 L25 average  jitter, degrees       14.343836---> cm  

%ocean model: chapron
%Size (m):       84.8528
%Resolution:      0.100000
%U10 (m/s):       4.00000
%fetch:       10000.0
%time_increment (s):    0.01000000
%Scaling:      0.200000  
%\end{verbatim} }
%\begin{figure}[b!] 
%%\hspace{2.5cm} 
%\epsfxsize=150mm 
%\epsffile{oldies/Sim-size=849pix^2,res=10cm,whrange=10cm.PIPhase.ps} 
%\caption{ \label{simul1} The results for a longer simulation: (Sim-size=849pixsq2,res=10cm,whrange=10cm) } 
%\end{figure} 

 \section{\sc speckles}
As mentioned above, this code has been developed to study the use of Geometric Optics for the study of ocean surface reflections in L-band.  It will also be used to study specular point statistics, using both Gaussian and, eventually,  non-Gaussian models.  Taking a Gaussian ocean model, we determine the specular points distribution and compute the reflected field using geometrical optics. We are to compare this to straight integration of Fresnel integral. Then the analysis will be  compared with  a non-linear ocean model.

\subsection{Specular point determination} 

We consider the incident radiation as a plane wave normal to a Gaussian surface and a receiver at $(x_r, y_r, z_r)$. The point $(x,y,\zeta)$ of the surface is considered to be a specular point if the slopes  follow rather clear geometrical conditions (see for instance [D.E.Freund 1997]):
\beq
%{d\zeta \over dx } = { { {x-x_r \over u_r} + {x-x_s \over u_s} } \over { {z_r-\zeta \over u_r} + {z_s-\zeta \over u_s} } }
{d\zeta \over dx } = {{x-x_r} \over {z_r-\zeta}}
\eeq
\beq
%{d\zeta \over dy } = { { {y-y_r \over u_r} + {y-y_s \over u_s} } \over { {z_r-\zeta \over u_r} + {z_s-\zeta \over u_s} } }
{d\zeta \over dy } = {{y-y_r} \over {z_r-\zeta}}.
\eeq

Our approach consists in the detection of slope variations around the expected specular slopes defined above. We compute at each point P of the surface, and
for both directions $x$ and $y$, the expected specular slope S---i.e. the slope needed to reflect the incident wave to the receiver---and the slopes of the facets located just
before (slope S[before]) and after (slope S[after]) the point P. The condition for P to be a speckle is that (S[before]-S) and (S -S[after]) must be of the same sign. This
determination is attractive because we doesn't need to define any tolerance on the specular slopes. We are neglecting saddle point by this approch (saddle points are not well defined in discrete spaces).
%It is obvious that the number of specular points depends on the "tolerance" applied to the slopes. The latter depends on the length of the receiver antenna and the distance $u_r$. We consider in our simulations an antenna length of 2 cm.

Then an analysis is made on the radius of curvature at the specular point. The Kirchhoff approximation leads to a condition on the radius of curvature at the specular facet, 
\beq
\sqrt{q\sqrt{r_1r_2}} >10.
\eeq
The product of the principal radii of curvature can be computed using  the derivatives of $\zeta(r)$, see [Barrick, 1968]:
$$
|r_1r_2| = { \left( 1 + \zeta_x^2 + \zeta_y^2\right)^2  \over \left|\zeta_{xx}\zeta_{yy} - \zeta_{xy}^2\right|  } 
$$

The following Figures (\ref{mylabel1} to \ref{mylabel3}) present the specular point position (circles) for three different times on the moving surface. The receiver is located at 8 meters above the surface, which has a 5 m by 5 m dimension and 1 cm of resolution. The stars correspond to the specular points that are under the curvature condition. The specular distribution is greatly modified within 10 ms.

\begin{figure}[h!] 
\hspace{2.5cm} 
\epsfxsize=90mm 
\epsffile{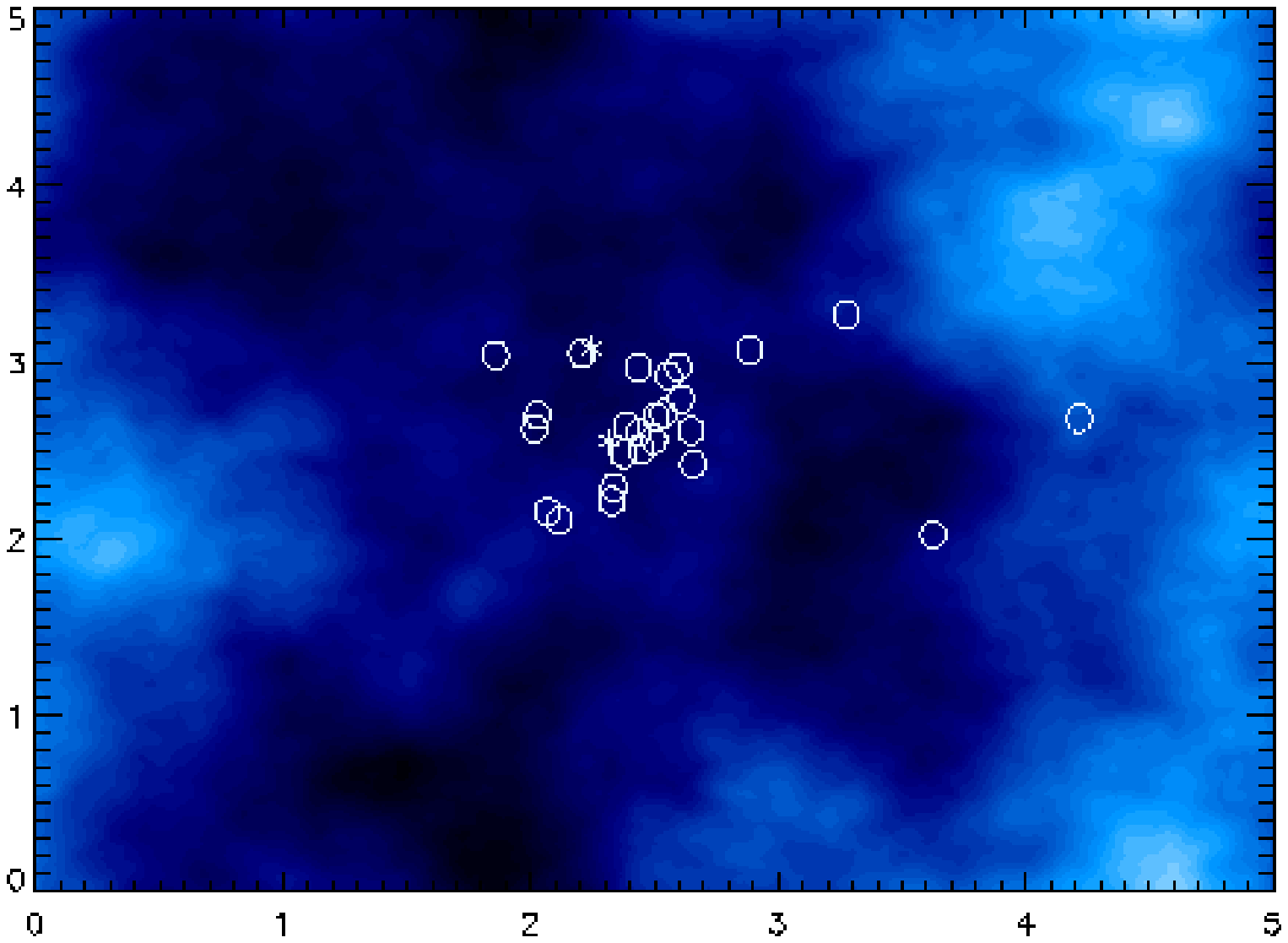} 
\caption{ \label{mylabel1} Specular points position for Time = 0 s} 
\end{figure} 

\begin{figure}[h!] 
\hspace{2.5cm} 
\epsfxsize=90mm 
\epsffile{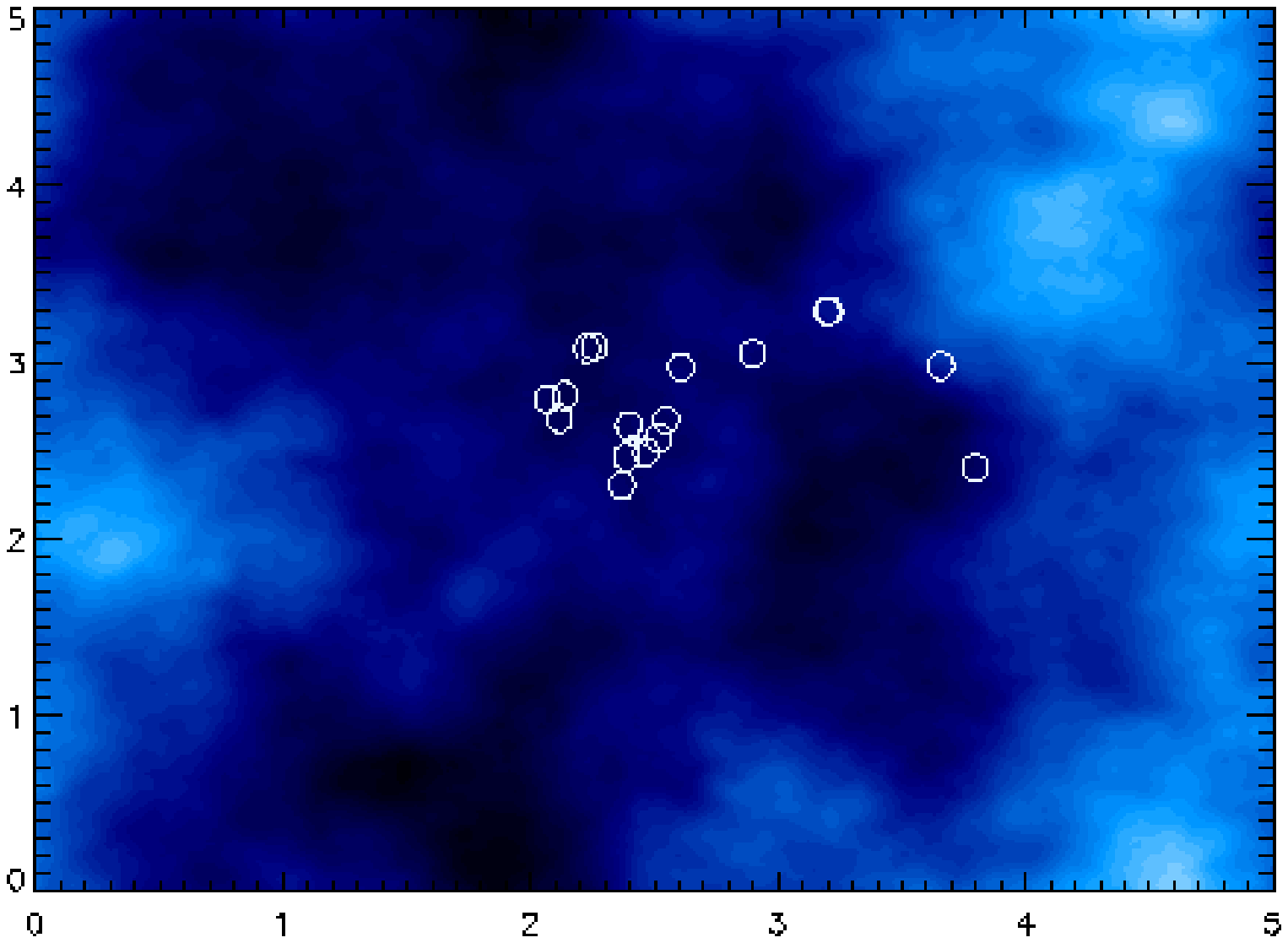} 
\caption{ \label{mylabel2} Specular points position for Time = 10 ms} 
\end{figure} 

\begin{figure}[h!] 
\hspace{2.5cm} 
\epsfxsize=90mm 
\epsffile{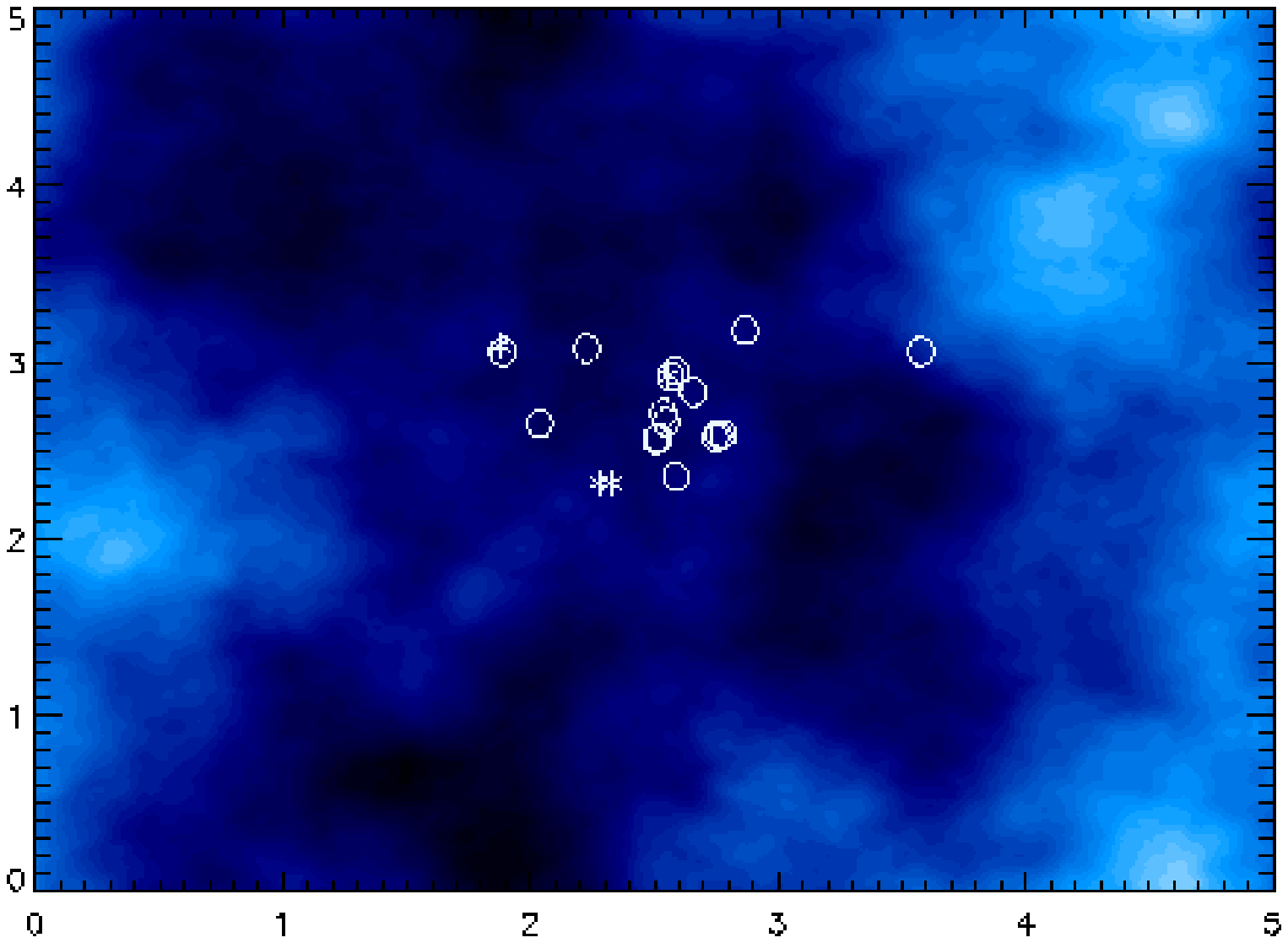} 
\caption{ \label{mylabel3} Specular points position for Time = 20 ms} 
\end{figure}

\subsection{Field computation}
The idea is to compute the Fresnel field with the contribution of scatterers only. We integrate on the vicinity of each mirror.  The result is then compared to the straight Fresnel computation. To date  these comparisons have not been very successful. It appears that in order for the Geometric Optics approach to work at this level (not after substantial averaging) we need to impose more stringent frequency or surface scale characteristics and/or take into account higher order effects. If the main contribution is from specular points,  as we suspect, the key may be to calculate the field from each scatterer to higher order than the stationary phase approximation.  Work in this area is planned for the future.  %We are now focusing on comparing specular point statistics to field. This work will be reported elsewhere (GNSS-OPPSCAT TN3230).
  
%%%%%%%%%%%%%%%%%%%%%%%%%%%%%%%%%%%%%%%%%%%%%%%%%%%%%%%%%%%%%%%%
\chapter{Coherence and Structure Functions}
\section{The coherence and  structure functions, and the coherence time}
Another question that needs to be answered  is how  the sea surface motion and varying geometry affect the coherence of the reflected signal.   There are several models for the sea spectrum, and these can be used to try to extract this information. Consider the  toy model of one-dimensional  scattering, with static, lined up, receiver and transmitter, with the scatterer (a single facet) also in line but now moving up and down with average speed  $v$.  It is easy to see that the  correlation of the direct and the  scattered signal is  zero unless  the coherent integration time is less than the characteristic time of the surface at GPS wavelengths.  The {\em  coherent integration time} is roughly that time for which the RMS phase error is 1 radian \cite{thomson}.  More precisely, let  the {\em coherence function}\index{coherence function}  be defined by
 \beq \label{cohfcn}
c(T)=\left| {1\over T}\int_0^T e^{i\phi(t)}dt \right|,	
\eeq
where $\phi(t)$  is the error phase.  Then, the coherence time is the time $T$  it takes for $\langle C(T)^2 \rangle$ to  
drop to 0.5---say \cite{thomson}.  Note that  for long times the coherence function  can also be related to  the magnitude of the zero-frequency spectral component of the (windowed)  normalized field associated to the error phase (that is, in our simulations, the Fourier transform of the normalized field).  We will  get back to this point below. 

The coherence time of the GPS signal is about 2--3 hours. What about the reflected signal? If the WAF zone and the sea surface are fixed, we should expect the same. If we include time dependence (and we ignore all motions), we can rewrite the equation on the statistics of the signal by
\beq
U(t)=r e^{i\psi} =\sum_j   e^{i\phi_j -i\omega t}=  e^{-i\omega t} \sum_j   e^{i\phi_j} ,
\eeq
so the resulting field is as coherent as the incident one.
 But in real situations  neither of this is the case. Both the ocean surface and the WAF patch  are moving (although this second effect can be corrected in principle by the use of SAR techniques).   The characteristic time of the surface can roughly be defined as the time it takes the surface to move a wavelength---in the present case $\lambda/v$. This thought experiment is useful to understand the limitation in coherent integration time  for a moving surface like the ocean.  An  advantage of using a synthetic wavelength using  two GNSS  frequencies is that this coherent integration  time should  be  longer. We will verify in a moment that  this is the case.

An associated concept to the coherence function is the {\em structure function} of the phase. This is defined by
\beq
s(T)=\langle \left( \phi(t+T)-\phi(t)\right)^2 \rangle_t .
\eeq
It gives us a measure of how tied up are two points in the phase as we vary their temporal separation. 
If this phase drift can be approximated by a random walk stochastic process \cite{Herring90},
\beq
s(T)=d\cdot T
\eeq
 we can use the associated ``Geophysically Induced Phase drift''  (GIP)  rate $\sqrt{d}$ as a geophysical parameter if it correlates well to U10 in our models. This parameters has units of cycle per square root second.  We will check this below (but let us anticipate that it does). 

\begin{figure}[h!]
\epsfxsize=145mm
\epsffile{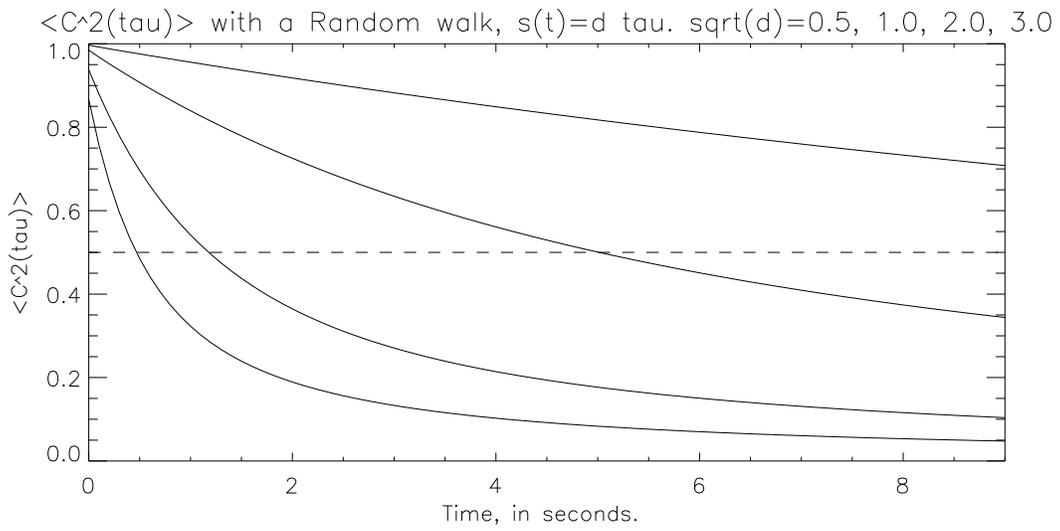}
\caption{ \label{c2}  $\langle C(T)^2 \rangle$ for a random walk with drift rates of $\sqrt{d}=$0.5, 1.0, 2.0 and 3.0 cycles per square root second.     }
\end{figure}

The coherence time  and the structure function are related. In fact, 
\beq
\langle C(T)^2\rangle={1\over T^2} \int_0^T \int_0^T \langle e^{i(\phi(t)-\phi(t'))} \rangle dt\, dt'. 
\eeq
 If we assume  Gaussian statistics for the phase fluctuations \cite{thomson} we can simplify things considerably, % and random walk stochastics, 
\bea 
\langle C(T)^2\rangle&=& {1\over T^2} \int_0^T \int_0^T e^{-s(t'-t)/2} \, dt\, dt'.\\ 
           %         &=& {1\over T^2} \int_0^T \int_0^T e^{-d \cdot (t'-t)^2/2} \, dt\, dt' \\
                    &=&  {2\over T} \int_0^T \left( 1-{\tau\over T}\right) e^{-s(\tau)/2} \, d\tau  .
\eea
For a random walk process, then, 
\beq
\langle C(T)^2\rangle = {2\over T} \int_0^T \left( 1-{\tau\over T}\right) e^{-d\cdot \tau/2} \, d\tau  .
\eeq
See Figure \ref{c2} for a graph of this function for various drift rates.

\begin{figure}[b!] 
\hspace{.5cm} 
\epsfxsize=130mm 
\epsffile{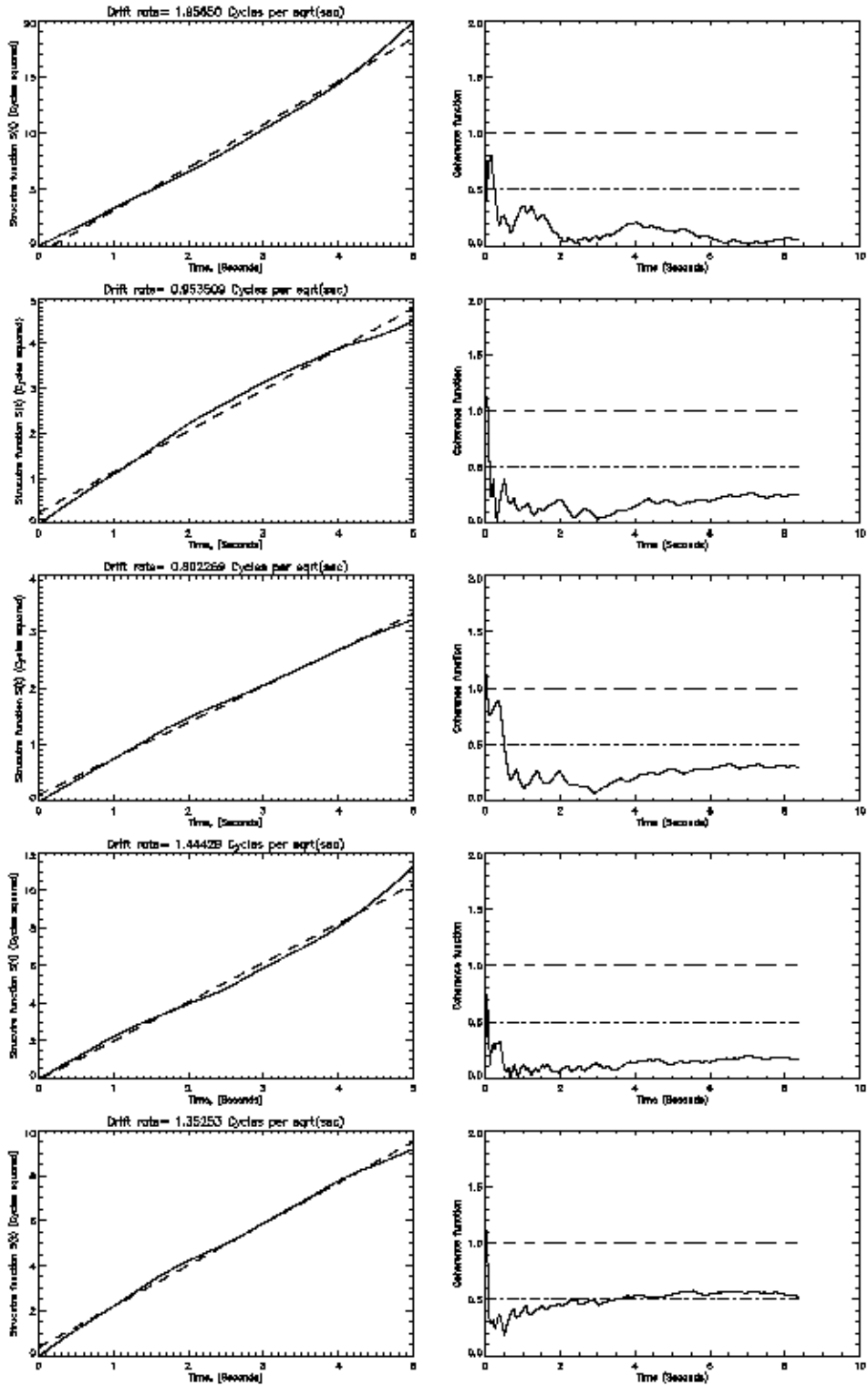} 
\caption{ \label{coherence100.ps} This is a plot with the structure  function (left) and the coherence function for the case U10=1 m/s. From top to bottom, L1, L2, L5 and L12, L25 fields. The drift rates are 1.9, 0.9, 0.8, 1.4, 1.3 cycles per square root second. Simulation size is 19 m side, 10 cm resolution, and height standard deviation is 3 cm.  } 
\end{figure}

\begin{figure}[b!] 
\hspace{.5cm} 
\epsfxsize=150mm 
\epsffile{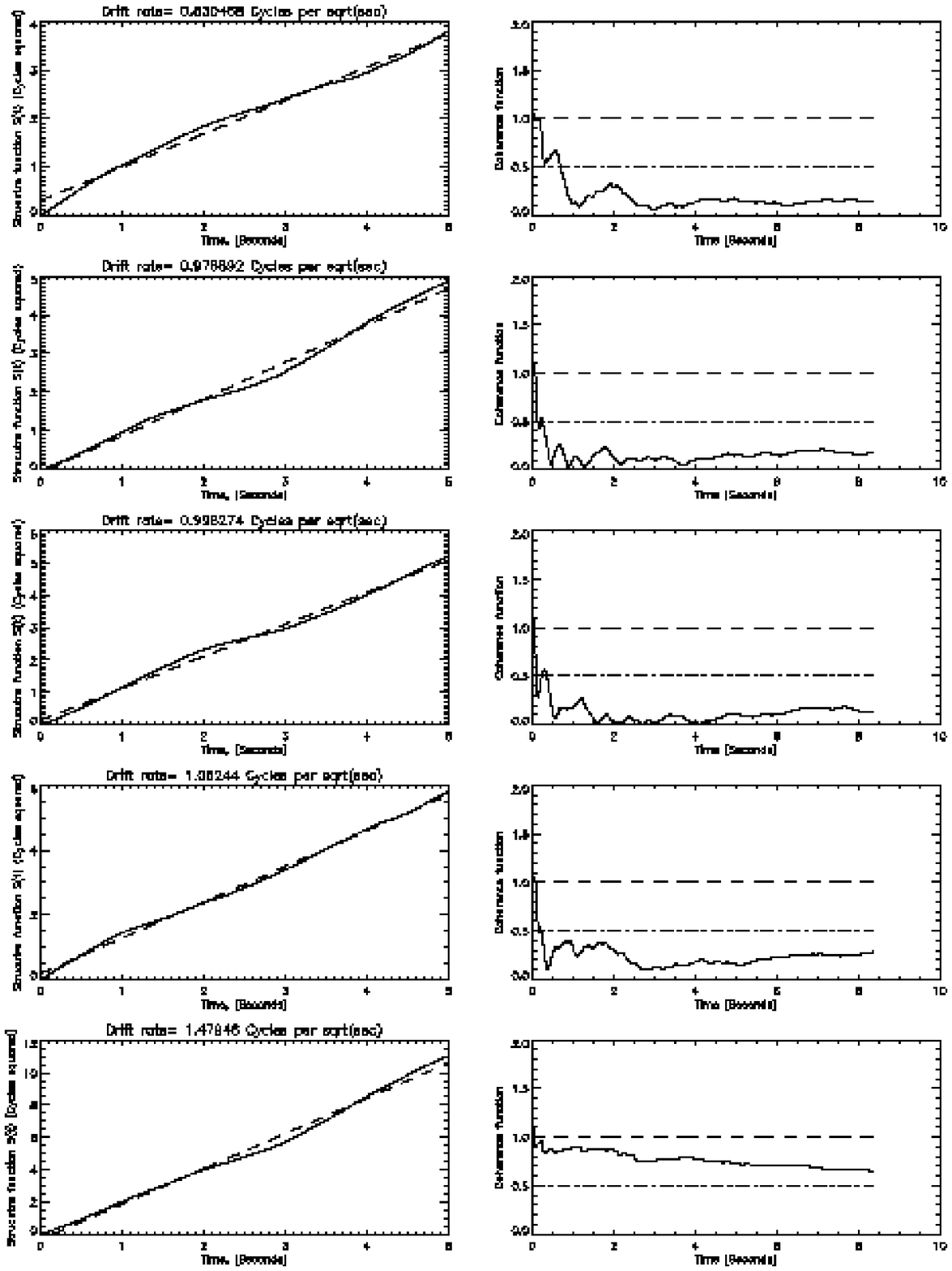} 
\caption{ \label{coherence300.ps} This is a plot with the structure  function (left) and the coherence function for the case U10=3 m/s. From top to bottom, L1, L2, L5 and L12, L25 fields. The drift rates are 0.8, 1.0,  1.0, 1.0, 1.5 cycles per square root second.  Simulation size is 19 m side, 10 cm resolution, and height standard deviation is 8 cm, receiver height 8 m. } 
\end{figure}

\begin{figure}[b!] 
\hspace{.5cm} 
\epsfxsize=150mm 
\epsffile{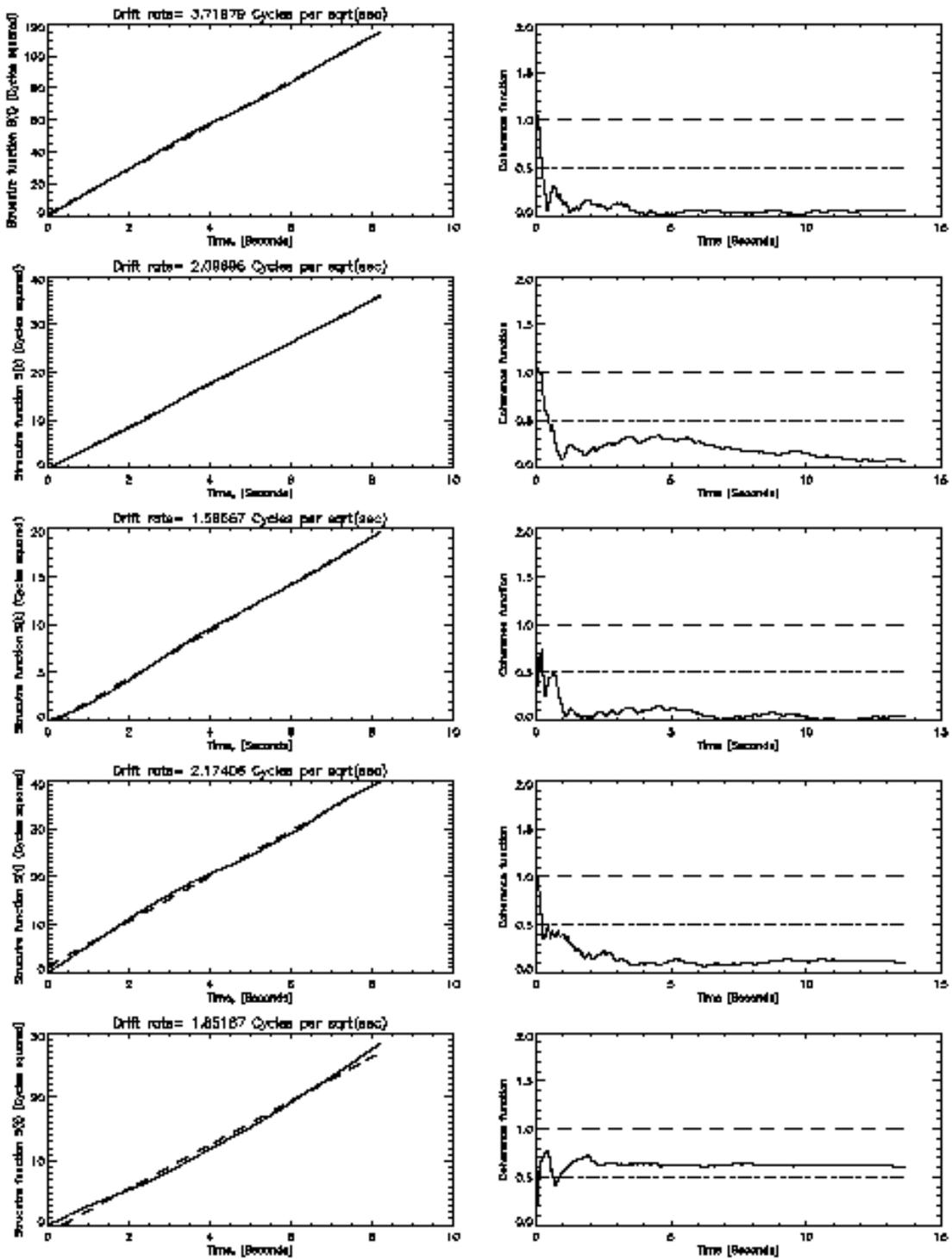} 
\caption{ \label{coherence400.ps} This is a plot with the structure  function (left) and the coherence function for the case U10=4 m/s. From top to bottom, L1, L2, L5 and L12, L25 fields. The drift rates are 3.7, 2.1,  1.6, 2.1, 1.8 cycles per square root second.  Simulation size is 19 m side, 10 cm resolution, and height standard deviation is 12 cm, receiver height 8 m.  See next Figure for further analysis of this particular simulation. } 
\end{figure}

\begin{figure}[h!] 
\hspace{.5cm} 
\epsfxsize=150mm 
\epsffile{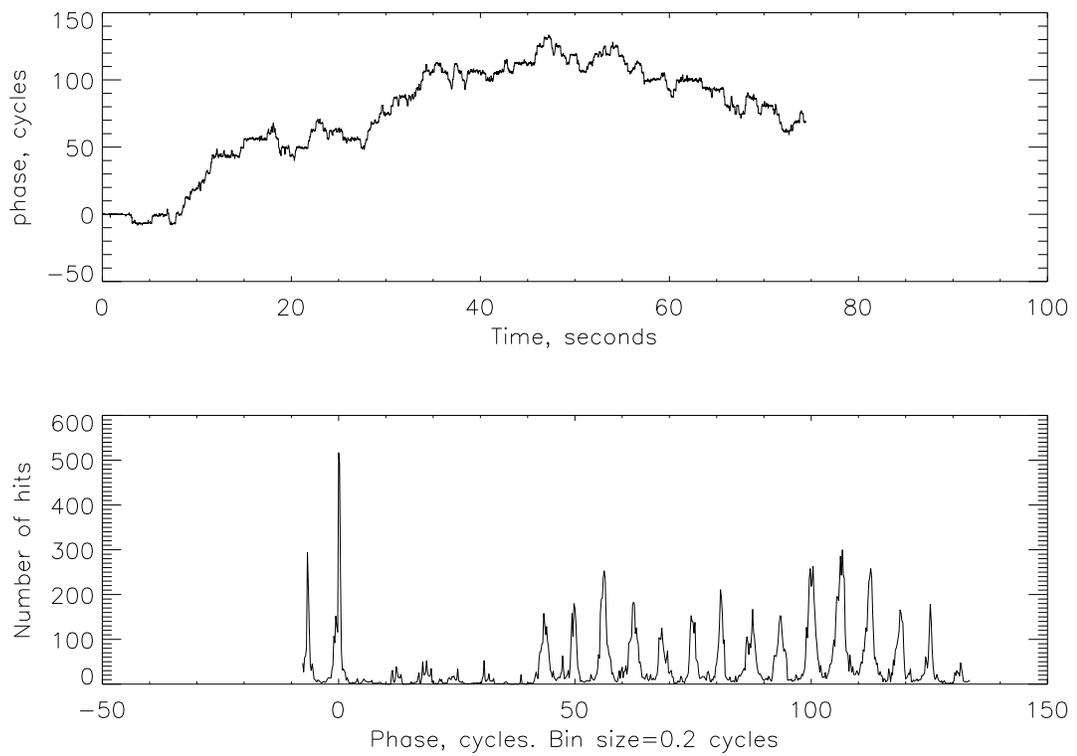} 
\caption{ \label{histo.ps} In these two plots we depict the phase and phase histogram for the L25 field in the previous figure. The highest peak corresponds to the initial zero phase. } 
\end{figure}

Note that in this model $\langle C(T)^2 \rangle$ decreases to zero as time increases. This fact alerts us that there is something amiss: we see in our simulations large phase excursions (certainly much larger than a cycle), but there is coherence left at large times---see Figures \ref{coherence_438pix2,r=10cm,U10=600cmpers.ps}, \ref{coherence100.ps},   \ref{coherence300.ps} and \ref{coherence400.ps}.   How is this possible? The Gaussian model for  phase fluctuations does not really apply to our situation, however. Gaussian means that phase excursions probabilities are bell-shaped, of course. Are they?  In our simulations the phase of the reflected signal gathers about a point in the complex plane---the {\em field}  does seem to have a Gaussian distribution about an average point.  The cumulative phase does not have to have such a distribution, however. In fact, it could conceivably  indulge in arbitrarily large excursions while spending more time in that average field point: this would still result in a Gaussian field distribution.  However, the probability distribution for such a phase history would not be a Gaussian  distribution (centered about 0, say). It would consist of a series of Gaussian-like humps spaced one cycle apart, with the one about zero the largest.  In this situation we could well  have a very long coherence time (since the phase modulo 2$\pi$ could  be a rather narrow  Gaussian distribution  about a point) while the structure function may show high drift. This type of behavior is evident in our simulations when plotting the field points (see Figure \ref{194pix2,r=10cm,U10=400fieldpoints.ps}). Note that the interferometric signal has a stronger average field value.   See Figure \ref{histo.ps} for a histogram of the L25 cumulative phase illustrating the multiple-hump distribution that we alluded to.

\begin{figure}[h!] 
\hspace{.5cm} 
\epsfxsize=150mm 
\epsffile{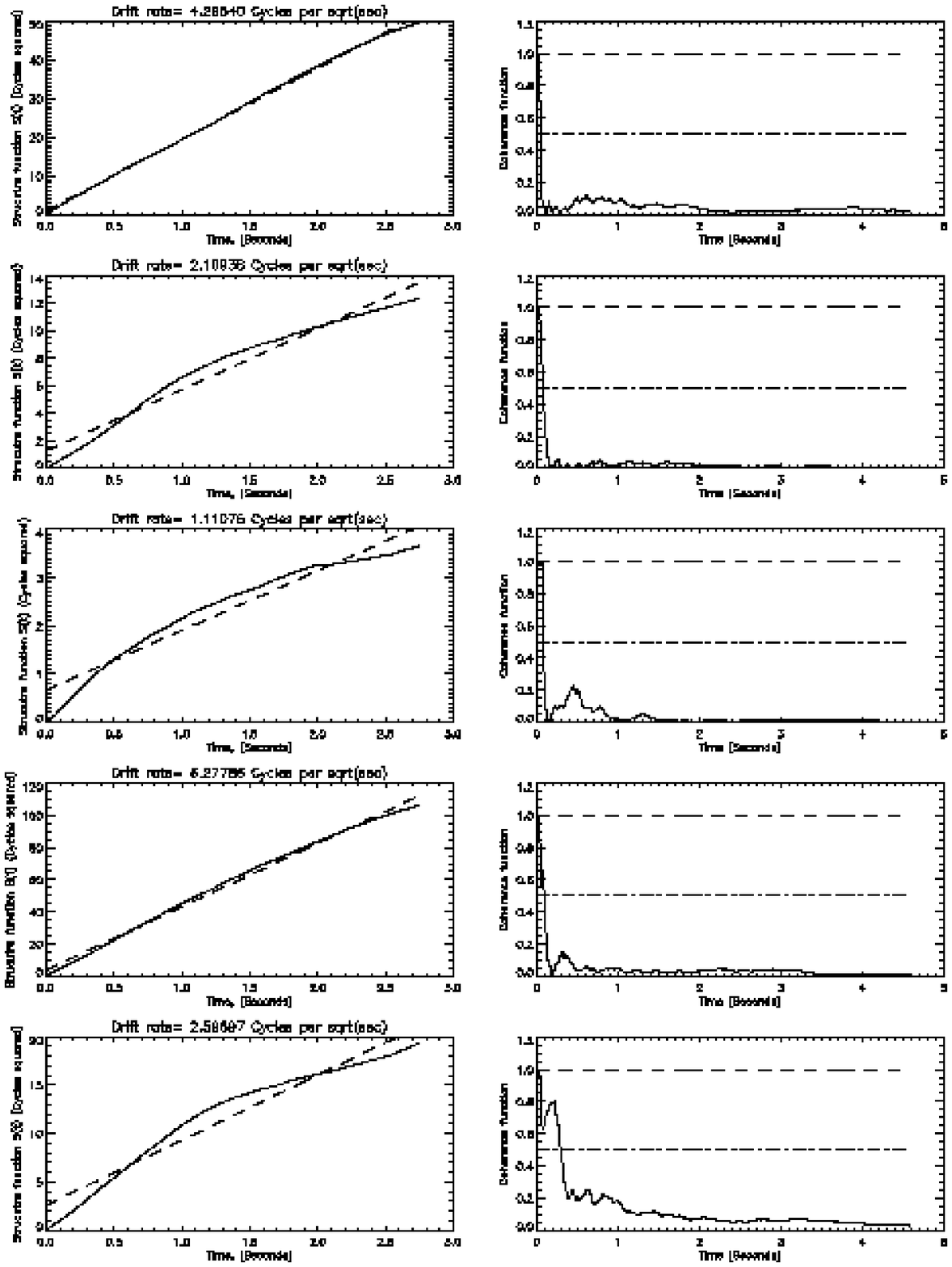} 
\caption{ \label{coherence_438pix2,r=10cm,U10=600cmpers.ps} This is a plot with the structure  function (left) and the coherence function for the case U10=6 m/s, 43 m side 10 cm resolution simulation at 8 m height.  This is  0.4 meters height standard deviation. From top to bottom, L1, L2, L5 and L12, L25 fields. The drift rates are 4.2, 2.1, 1.1, 6.2, 2.5 cycles per square root second. Coherence times 0.04, 0.06, 0.08, 0.09 and 0.3 seconds. } 
\end{figure}

%\clearpage
\begin{figure}[h!] 
\hspace{3.5cm} 
\epsfxsize=200mm 
\epsffile{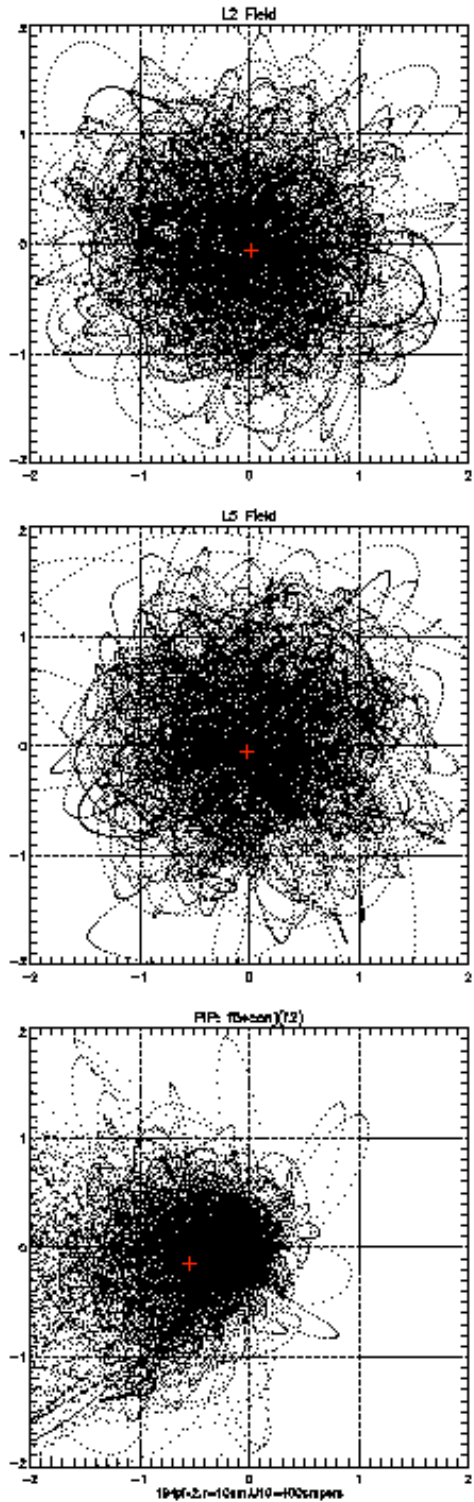} 
\caption{ \label{194pix2,r=10cm,U10=400fieldpoints.ps} Refer to the previous figure. This is a plot of the field in the complex plane for L$_2$, L$_5$ and L$_{25}$, and it serves to illuminate the difference between the structure function and the coherence function.  The field mean for L$_2$ is (-0.06,-0.1), for L$_5$ is (-0.1,-0.1), and for L$_{25}$  (-0.6,-0.20), much larger. The standard deviation is about 1 for all of them.  In this 19 meter simulation, U10=4 m/s, height standard deviation was 12 cm, receiver height 8 m. } 
\end{figure}

\clearpage

Notice that the above discussion is relevant to understand  the coherence properties  that were found in the Bridge Experiment reflected signal (integration times where of 10 ms, short enough for the static approximation to work in that geometry).  In longer time scales the accumulated phase will pick up a random component, but if there is an overall phase  change (due for instance, to a tide) it should  be distinguishable from the random part (which should not add nor subtract to the overall phase on the average).  
\section{Doppler spread of the reflected signal}
\begin{figure}[b!] 
\hspace{1.5cm} 
\epsfxsize=160mm 
\epsffile{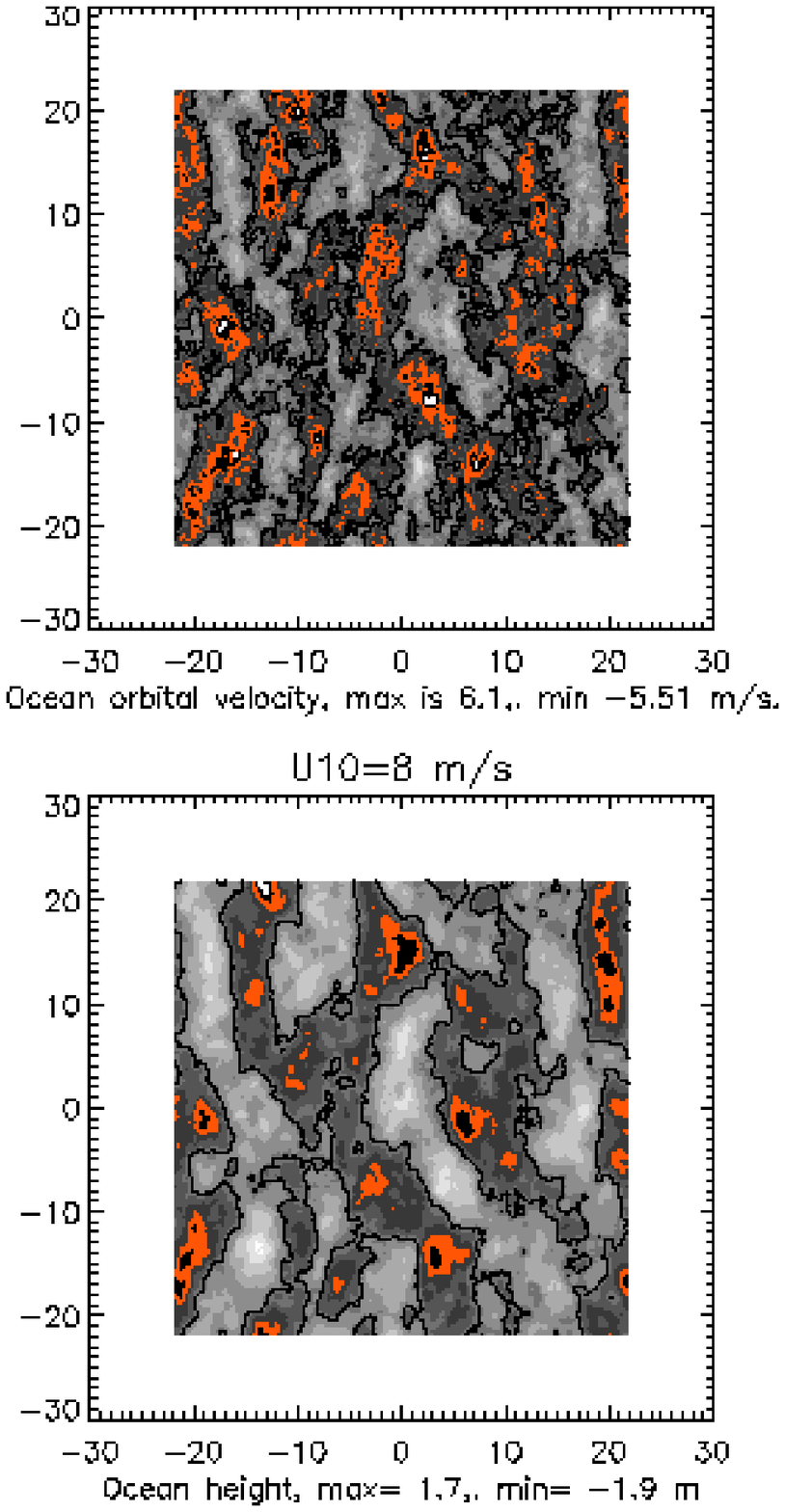} 
\caption{ \label{etaandveta.ps} The ocean surface and vertical velocity for a wind of $U10=8$ m/s, Elfouhaily spectrum. The contour denotes the zero value in each case. } 
\end{figure}

Another thought experiment is useful. Imagine that the surface height is varying as a function of time as
$$
\zeta(t)=\alpha t +A r(t),
$$
where $\alpha$ is a constant and the second term adds a random, noise-like displacement between $-A$ and $A$. We can think of this as the  effect of  a tide superimposed on the wave motion  on the sea.  The scattered phase will not be coherent if A is large enough, but the accumulated phase will certainly contain information about $\alpha$.  What happens if there are many scatterers, how fast is the residual phase varying?  This will give us a feel for the Phase Lock Loop (PLL) bandwidth required for tracking. It is not hard to see that the bandwidth required  will be less than 100 Hz (assuming that the sea surface vertical speed is not greater than 10 m/s), in a static geometry (or if a static-WAF is used). The largest frequency present in the signal cannot be greater  than the contribution due to the fastest moving  surface patch---the result is always the linear combination of such contributions, $
U(t) \sim r e^{i\psi(t)} =\sum_{j=1}^n   e^{i\phi_j(t)}$.
 Now, if the maximal orbital velocity for a wave is 10 m/s (see Figure \ref{etaandveta.ps}, this implies a Doppler of $2\times10/\lambda \approx 100$ Hz (remember that the signal is bouncing off the surface).  Thus, a sampling rate of at least 200 Hz is required to retrieve the phase of the reflected signal (even if the SNR is high enough).  In our simulations we have seen that it is safe to sample the signal at such rate.

\begin{figure}[h!]
%\hspace{2.5cm}
\epsfxsize=145mm
\epsffile{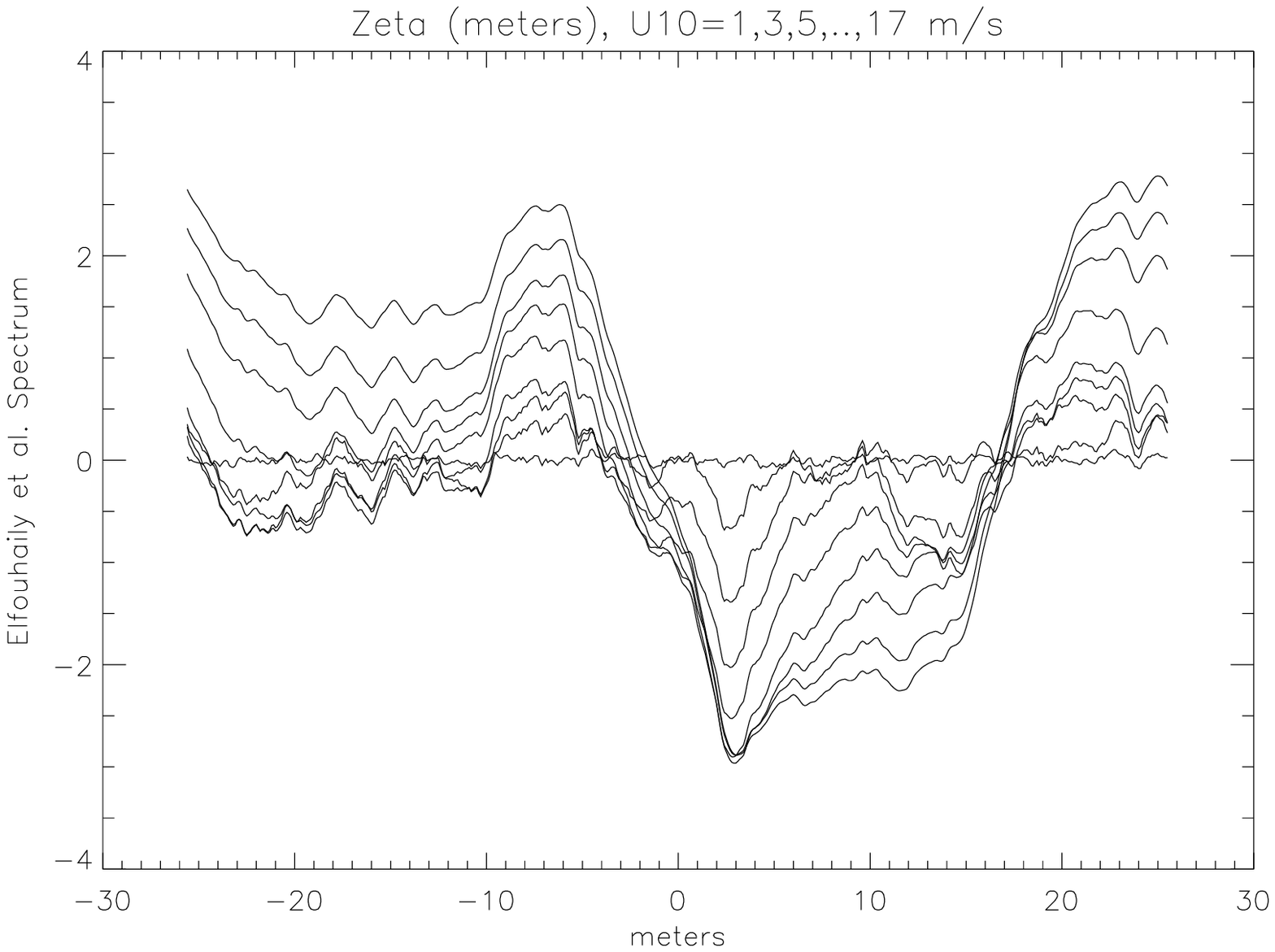}
\epsfxsize=145mm
\epsffile{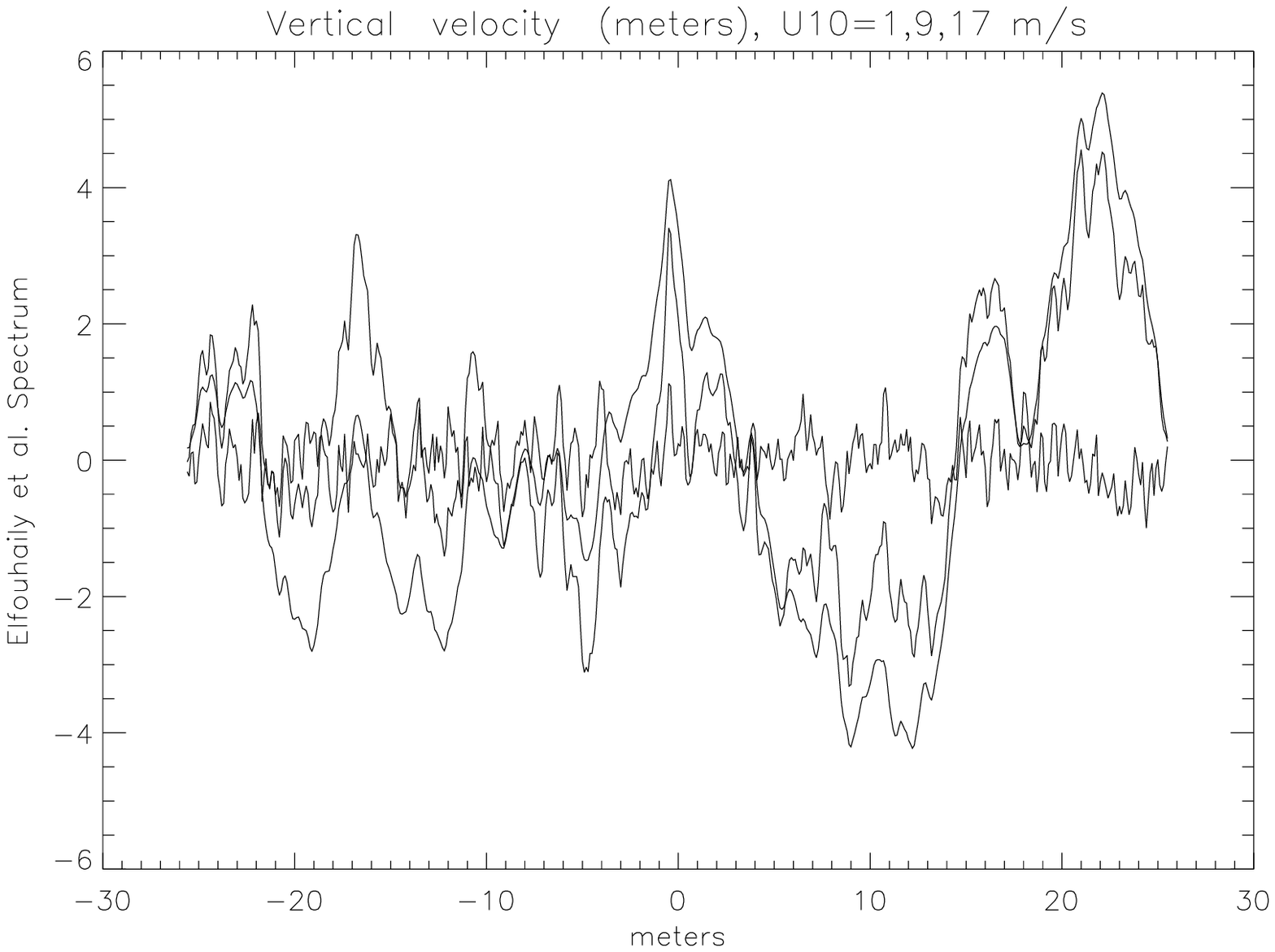}
 \caption{ \label{zeta} Ocean heights and vertical velocities using the Elfouhaily et al. spectrum for wind speeds of U10=1,3,5,..,17 m/s.    Note that due to limitations in the simulation size values for U10 above  12 m/s suffer from saturation.  }
\end{figure}

Here is another useful  question: given that a receiving LEO is moving at, say, 7 km/s, would it be possible to integrate coherently (correlate)  the direct and reflected signals for 1 ms, if the LEO antenna has a footprint of 10 meters? Note that this ``eye on the sea'' is moving very fast over the surface, and this is introducing a random, time-dependent phase to the signal. The answer is no. But if the footprint is enlarged enough, yes. This may be an important problem for the PIP, but there are ways through which it could be fixed. One obvious one is to use a WAF that sticks to a given surface patch, or that it moves slowly over the sea surface.  Some of these ideas are discussed within the context of the GNSS-OPPSCAT project (Report WP3320).
To get an idea of the induced Doppler, we can reason that the scattered field is just a sum of the type
$$
F= R(t)e^{i\Phi(t)}\approx \sum^N_{i=1} r_i(t) e^{i\phi_i(t)},
$$
which already looks like a Fourier sum if we approximate $r_i(t)$ as constant and the exponent by its first derivative.  Thus, we can guess that 
the maximum frequency in the spectrum is simply Doppler due to the fastest moving patch on the ocean. As an upper bound we can use 10 m/s---see Figure~\ref{zeta}. Multiplying times 2 and dividing by $\lambda$ we obtain a scattering Doppler of  about 100 Hz. 

To be a bit more precise, let us write
$$
F=\int_{A_{WAF}} e^{iq_z z} d^2x=\int_{A_{WAF}} e^{iq_z(z_0+\dot{z}t)} d^2x,
$$
using the simpler backscattering expression in the Fraunhofer zone.  Notice the restriction of the integral to the (static) WAF zone.  Let us now rewrite the transform of the field:
\bean
\tilde{F}(\omega)&=&\int Fe^{i\omega t} dt= \int e^{i q_z z_o}\int e^{i(\omega t-q_z\dot{z})}    dtdx\\
&\sim& \int_{A(q\dot{z}\equiv \omega)}  e^{-i q_z z_o} d^2x,
\eean
if we assume that vertical motions take over a ``long enough'' time compared to the time scale associated to the frequency we are interested in. This is saying that the spectral component at a given frequency is proportional to the area on the surface moving at the right speed to produce that frequency.

We can extend this reasoning to the moving-receiver  case.  
There are two limiting situations we can think about. One is the ``high gain'' situation. Let us first discuss this situation---we will get back to the low gain situation at the end.  Assume for now the ocean is frozen. In this situation the WAF zone is very small (due to antenna or processing gain) and coherence time is definitely influenced by the fact that as the receiver moves the contributing surface is changing. If the WAF zone is smaller than the correlation length of the ocean (certainly an extreme case, but never mind), the coherence time is limited  by $l/v$, where $v$ is the receiver speed and $l$ is the correlation length of the surface---assuming large enough height deviations.  To be more precise, 
if we use a replica with fixed Doppler and Delay, the ``eye on the sea'' will move over the ocean at a speed similar to the receiver's.  This will introduce additional  bandwidth in the carrier if the ocean is rough---even if is frozen.
This bandwidth will be mainly proportional to ocean surface rugosity, not to ocean surface motion. It can be very large (kHz), depending on the receiver speed and the WAF area size.  Following the above reasoning, and keeping in mind that the WAF zone is now a moving filter on the ocean surface,
\bean
F(t)&=&\int_{A_{WAF}(t)} e^{iq_z z(\vec{x})} d^2x= \int  \Xi^{WAF}(\vec{x}-\vec{v}t) \,  e^{iq_z z(\vec{x})} d^2x \\
&=& \int  \Xi^{WAF}(\vec{x}) \,  e^{iq_z z(\vec{x}+\vec{v}t)} d^2x \\   
 &=&  \int_{A_{WAF}}  \,  e^{iq_z z(\vec{x}+\vec{v}t)} d^2x 
%&=&  \int_{A_{WAF}} e^{iq_z [ z(\vec{x})+ (\nabla z\cdot \vec{v} + \partial_{t} z) t] } d^2x
\eean 
Now let us look at $\dot{F}(t)$:
\beq
\dot{F}(t) = \vec{v}\cdot \int_{A_{WAF}}  \, \vec{\nabla} \left(  e^{iq_z z(\vec{x}+\vec{v}t)}\right)  d^2x=   \vec{v} \cdot \int_{\partial A_{WAF}}  \, e^{iq_z z(\vec{x}+\vec{v}t)} \ \hat{n} \  dl
\eeq
where the last is an integral over the boundary of the WAF area and $\vec{n}$ is the normal to the boundary. The result illustrates the  ``edge effect'', i.e., that the change is just due to the change at the boundary of the WAF area. We have used a 2D version of Green's theorem in the last step. This equation says that the change in the field is proportional to the variation of the field contribution at the edges of the WAF area (properly mapped in the velocity direction). The faster the velocity and the larger the difference, the larger the rate of change of the field. To relate this to the spectrum, note that this will contribute a high frequency componene to the spectrum. Whether this is a relatively large or small contribution depends on the total field, which is proportional to the total area. Thus, for large WAF zones this high frequency effect contributes a small portion of the total spectrum. 

If we allow for a moving surface, $z=z(\vec{x},t)$ it is readily seen that the complete result is the sum of two distinct effects:
\beq
\dot{F}(t) =   \vec{v} \cdot \int_{\partial A_{WAF}}  \, e^{iq_z z(\vec{x}+\vec{v}t,t)} \ \hat{n} \  dl +  
\int  \Xi^{WAF}(\vec{x}-\vec{v}t) \, \partial_t  \left( e^{iq_z z(\vec{x},t)} \right) d^2x.
\eeq

Let us now return to 
\beq
F(t)=\int_{A_{WAF}(t)} e^{iq_z z(\vec{x})} d^2x= \int  \Xi^{WAF}(\vec{x}-\vec{v}t) \,  e^{iq_z z(\vec{x})} d^2x
\eeq \
and compute the Fourier transform:
\beq
\tilde{F}(\omega) = \int  \int_{-\infty}^\infty dt \ e^{i\omega t} \ \Xi^{WAF}(\vec{x}-\vec{v}t) \,  e^{iq_z z(\vec{x})} d^2x.
\eeq
To evaluate  this, let us assume for simplicity and without loss of generality  that $\Xi^{WAF}$ is a box of size $S$ moving along in the $x$ direction with velocity $v_x$:
\beq
\Xi^{WAF}(\vec{x}-\vec{v}t) = \Xi_x(x-v_xt) \cdot \Xi_y(y).
\eeq  
Then 
\beq
 \int_{-\infty}^\infty dt \ e^{i\omega t}\  \Xi^{WAF}(\vec{x}-\vec{v}t) = 2{ \Xi_y(y) e^{i\omega x/v_x} \over w} \sin  (\omega S/v_x). 
\eeq
Hence,
\beq
\tilde{F}(\omega) ={ 2 \sin  (\omega S/v_x ) \over \omega}  \int \Xi_y(y)\  e^{i\omega x/v_x} \ e^{iq_z z(\vec{x})} d^2x. 
\eeq
The characteristic time is roughly given by $S/v_x$, as can be seen from the multiplying sinc. Frequencies higher than $v_x/S$ are supressed by this factor. The power in frequencies smaller than this are  modulated by the horizontal roughness of  $\exp{iq_z z(\vec{x})}$ in the scale defined by $ v_x/\omega$  in the direction of motion: if the surface does not vary in that scale, there will be little power at the frequency $v_x/S$.

In this approximation we have implicitly assumed a very small WAF in relation to the geometry, since we have worked in the Fraunhofer approximation. This means that all the points in the reflecting patch have the same geometry and generate the same geometrically induced Doppler. In recent experimental conditions this is hardly the case, and the Doppler of the reflected signal is dominated by the Doppler amplitude of the WAF and glistening zones. If the glistening zone is large enough, then radiation will be received from the entire WAF. The Doppler span of the WAF zone determines then the Doppler spread of the received field. This is the basic mechanism for Delay-Doppler mapping and  the basis for the relationship betweeen waveforms and the characteristics of the sea surface.  Even if we restrict the signal  to the first chip in the C/A code, there will be quite a bit of geometric Doppler.  This second source of Doppler is in principle removable by many means: high antenna gain or processing gain. By any of these means we can reduce the patch to the Fraunhofer zone mantaining SNR. It will then become increasingly important to use SAR techniques to focus and eliminate the edge effect we just discussed.   We plan to revisit these calculations extending them to  the Fresnel zone in future work. 

The  ultimate  limit to the Doppler width and therefore to the coherent integration time of the signal is ocean motion--that cannot be anticipated and compensated for. 

The size of the first Fresnel zone for a receiver at 350 km is about $2\sqrt{2h\lambda}$ or about 1 km. In this narrow area, the geometric Doppler spans about 200 Hz edge to edge (the receiver looking down and is moving along at 7 km/s) and the corresponding integration time is 5 ms.  This Doppler is of the same order of magnitude of  ocean induced spread. Integration times longer than 10 ms will certainly begin to be sensitive to ocean motion. The edge effect we discussed earlier will also start to play a role at longer  integration times : for a 1 km WAF (roughly corresponding to 5 ms integration time) the corresponding frequency is of 7 Hz--- still too small to make spotlight processing necessary.   This may be the best way to retrieve ocean  induced Doppler spread. See Table-\ref{tab} for a summary of these ideas.  

\begin{table}
\centering
\begin{tabular}{|c|c|c|c|} \hline 
$T_i (ms)$ & Ocean motion (Hz) & Geometric (Hz)  & Edge (Hz)\\ \hline \hline
1&     200  &  1000& 1  \\ \hline
5&    200 & 200 & 7  \\ \hline
10&   200 & 100 & 14 \\ \hline
20&    200& 50 &  28 \\ \hline
50    &200 &  20 &70  \\ \hline 
\hline
\end{tabular}
\caption{Back of the envelope calculations for Doppler spread induced by ocean motion, geometry and edge effects. Note that above 20 ms edge effects force the use of focusing.}
\label{tab}
\end{table}

%Let 
%\beq
%\Phi(t)=q_z z(\vec{x}+\vec{v}t).
%\eeq
%Then,
%\beq
%\tilde{F}(\omega) = \int_{A_{WAF}}    e^{i\Phi(t)-i \omega t} dt d^2x 
%\eeq
%In terms of the ``instantaneous frequency'', we can read off
%$$
%\tilde{F}(\omega) \sim \int_{A(q_z(  \nabla z\cdot \vec{v} + \partial_{t} z )\equiv \omega)}  e^{-i q_z z_o} d^2x .
%$$
%There is now much  more Doppler in the reflected signal, just as we expected. This depends, again, on geophysical parameters (ocean velocity and rugosity).  For a LEO moving at 7 km/s, for instance, $\nabla z\cdot \vec{v}$ can be  as large as $\sqrt{0.035}*7000 *2* \lambda = 13$ KHz. For an aircraft moving at 180 m/s and U10=5 m/s, we get a maximal Doppler of 180 Hz + 10 Hz (including Doppler due to vertical motion of about 1 m/s). Note that the large Doppler we just found can be corrected by the use of a ``matched filter'' replica, static or slowly drifting over  the surface.  
%\begin{figure}[h!]
%%\hspace{2.5cm}
%\epsfxsize=100mm
%\rotatebox{-90}{\epsffile{seamodels.ps}}
% \caption{ \label{elf} Mean square slopes in the Elfouhaily spectral model.}
%\end{figure}
We  have two different new ways to extract sea-state information from the Carrier bandwidth, by playing with the eye location. One the one hand, we can use a static WAF to measure vertical velocities---this will entail using SAR techniques from moving platforms. On the other, we can measure surface roughness  by letting the WAF drift over the surface at a speed of choice.  This is just a matter of tuning the matched filter appropriately, and   it could even be done from a static platform.  % The physical picture to keep in mind is that of a moving searchlight. We are moving a coherent searchlight over a surface, and learning about its topography by keeping track of the phase. More on this in the next section, which falls in the same logical framework as this one.es.

  %To see this, write
%$$
%U(t)=r e^{i\psi(t)} =\sum_{j=1}^n   e^{i\phi_j(t)},
%$$
%and assuming $r$ to be almost constant (Central Limit Theorem)
%$$
%|\dot{\Psi}(t)|= \left| {\sum_{j=1}^n \dot{\phi}_j(t) e^{i\phi_j(t)} \over \sum _{j=1}^n   e^{i\phi_j(t)}}
%\right| \leq { n | \dot{\phi}_j(t)|_{max} \over r} =  { n | \dot{\phi}_j(t)|_{max} \over \sqrt{n^2 e^{- \sigma^2} +n(1-e^{- \sigma^2})}}  
%.  
%$$
%In the rough limit we have
%$$
%|\dot{\psi}(t)| \leq { \sqrt{n} | \dot{\phi}_j(t)|_{max}}
%$$
%If we think of the case with 10000 scatterers (pessimistic!), each contributing about an Hz,  we have about 100 Hz bandwidth in the resulting signal. Note that an immediate consequence of this analysis is that the bandwidth of the reflected signal ia a potentially  rich source of geophysical information.

%For  a fully developed sea, a wind  of 10 m/s needs to blow with a duration of 18 hours over about 320 km of ocean, and under these conditions SWH ($\zeta_{1/3}$) will be of about 2 meters and wave period will be about 7.5 s (see \cite{Ruffini99} for details). This means that vertical speeds are about   0.5 meter per second. It thus takes about 0.5 seconds to displace the sea surface a wavelength at GPS frequencies.  This is 

%\begin{figure}[b!] 
%\hspace{.5cm} 
%\epsfxsize=150mm 
%\epsffile{coherence1500.ps} 
%\caption{ \label{coherence1500.ps} This is a plot with the structure  function (left) and the coherence function for the case U10=15 m/s. From top to bottom, L1, L2, L5 and L12, L25 fields. The drift rates are 5.2, 3.6, 4.1, 3.5, 2.1 cycles per square root second.  } 
%\end{figure} 

\chapter{Statisitical Properties of the Reflected Fields}
\section{Analysis of field correlations}
In this section we address the question of the correlation between the fields at different frequencies (say $L_1$ and $L_2$ and the future $L_5$) for different sea conditions. An approach to this problem is to rewrite the Fresnel integral using the ``zone'' concept. That is, we classify areas in the surface according to their distance to the receiver. We assume we are in the Fraunhofer-emitter zone, so the distance to the emitter plays no role in this discussion. The idea is then to rewrite the field as a sum of field contributions from equal delay ($r$)zones:
$$
U_{q}=q\int_{r_{min}}^{r_{max}} e^{iq r} {\cal A}'(r) dr.
$$
This idea can be found in \cite{Berry72}. The function ${\cal A}'(r)$ is an ``areal'' density function that takes into account how much area contributes to each zone. It's exact expression is not of immediate concern. In the case of a flat surface it just becomes one.  In order to extend the integration over the whole $r$-axis, let us simply extend the definition of ${\cal A}'(r)$ to be zero outside ${r_{min}}$ and $r_{max}$.   We then have,
$$
U_{q}=q\int_{-\infty}^\infty e^{iq r} {\cal A}'(r) dr.
$$
We can now read: the field is the Fourier transform of ${\cal A}'(r)$. And the question about the relationship of the field between frequencies becomes a question on the relationship between the Fourier components of this areal function at different frequencies. 
%Since we are looking at the Fourier transform of a ``pulse'' of size  $L={r_{min}}-r_{max}$, 
%we can already guess that if $\Delta  q \cdot L$ is small, then there will be correlation between the two frequencies. In generasl, 

Let us forget for now the meaning of the areal function, and just think in terms  of the Fourier transform of a function with limited support, since we are looking at the Fourier transform of a function with  support  $L\equiv {r_{min}}-r_{max}$.  The larger the support of the function, the more small-scale structure can 
be found in the Fourier transform---everything else the same. We are accostumed to thinking about this fact in the opposite domain: a function with  a lot of small time-scale structure will have a large bandwidth---a large support in the frequncy domain. 

Intuitively, the larger the support    two given nearby frequency  
components  will end up sampling  the pulse at more separated regions.  It is useful to think 
about two sine functions of slightly different frequencies, running side by side. As a rule of thumb, we expect that the correlation between these Fourier components will be sensitive to the 
correlation function of the (areal) function at a distance  $ (\Delta\lambda)*L/\lambda_{mean}$.  This is because $\Delta \lambda$ is the sampling distance 
difference gained per cycle, and $*L/\lambda_{mean}$ is the number of cycles available in a region of size $L$.
Therefore, before carrying out any calculation we can say that if the areal function  is correlated at the distance dictated by this  (maximal possible) 
frequency separation, then we will see correlation in the fields.  It helps to look at the separation between the equi-delay contours in the surface of 
integration. If this separation is smaller than the correlation length of the surface, then  high correlation between the fields is possible.
There is an extra consideration, however. We have been assuming that the separation between contours is constant, but this is not the case. That is,  we have been implicitly assuming that the contours are fixed on the surface, and all we have considered is the correlation between this contours. But in fact, as the ocean moves, the contours move, especially with large seas. This introduces an additional source of decorrelation, which gets worse with larger seas.
The time-correlation between the fields is  given by 
\bean \label{beast}
\langle  U_{q_1} U_{q_2}^* \rangle_t &=&q_1q_2\int \langle {\cal A}_1'(r) {\cal A}_2'(r') \rangle_t  e^{i q_1 r -iq_2 r'} dr dr'.
\eean
This is a difficult beast to deal with, so we will change strategy.

The correlation between the fields can also be calculated in another manner.   It will be useful to approximate things entirely in the Fraunhofer zone, i.e., with both transmitter and receiver in the far field, assuming Gaussian statistics. 
The field in the nadir case is given, in 1-D and up to a constant, by
\beq
U=q\int_A e^{iqz(x)} dx. 
\eeq 
Now, 
\beq
|U|^2= q^2\int_A\int_A e^{iq\left(z(x)-z(x')\right)} dxdx',
\eeq
and
\beq
\langle |U|^2 \rangle = q^2\int_A\int_A \langle e^{iq\left(z(x)-z(x')\right)}\rangle dxdx'.
\eeq
Now, if we assume Gaussian statistics, with
\beq
P(z(x)-z(x'))= {1\over 2 \pi \sigma^2 \sqrt{1-\rho^2(x-x')}} \exp\left[ { - {z(x)^2-2\rho(x-x')z(x) z(x') +z^2(x') \over 2\sigma^2(1-\rho^2(x-x'))}}\right] ,
\eeq
we can carry out these calculations explicitly. Here $\sigma$ is the height standard deviation from the (zero) mean, and $\rho$ is the correlation function of the surface. It is healthy to keep in mind that in the gaussian case, there are the only parameters, together with the frequencies, that can appear in the final expressions. If we further assume a Gaussian correlation function with correlation length $l$, the answer to any question must be expressed in terms of dimensionally meaningful expressions contantaining, $\sigma$, $l$, $\lambda_1$ and $\lambda_2$. 
\begin{figure}[b!] 
\hspace{1.5cm} 
\epsfxsize=120mm 
\epsffile{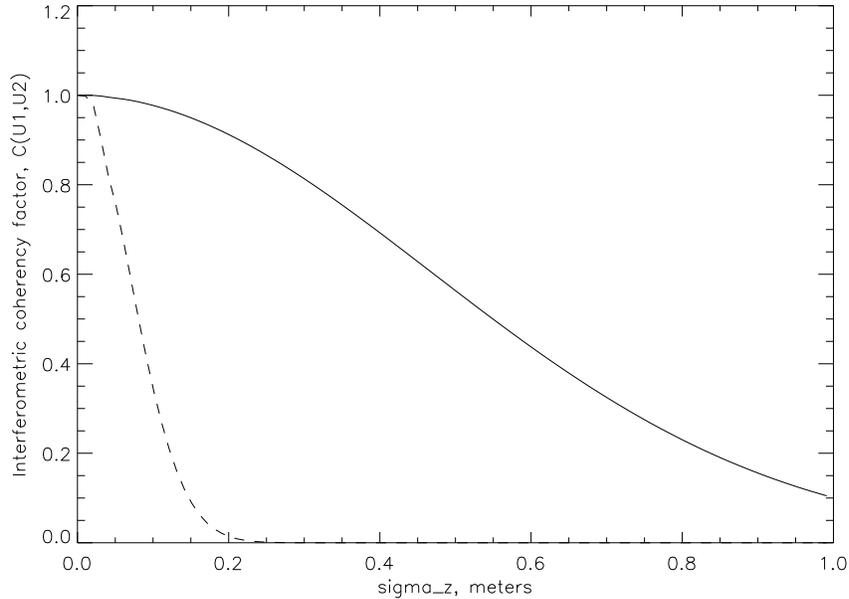} 
\caption{ \label{coherency} The coherency factor $C(U_1,U_2)$ (dashed) and $C(U_2,U_5)$ for the case $l=5$ and $L=20$. There is not much sensitivity in either case to $L$ or $l$.   } 
\end{figure}

For the case at hand, the key result is (\cite{Beckmann}, p. 190)
\beq\label {average}
\langle e^{iq_1z(x)+q_2z(x')}\rangle = \exp \left[ -{1\over2}\sigma^2(q_1^2+2\rho(x-x')q_1q_2+q_2^2)\right ].
\eeq
This implies, for instance,
\beq
\langle U_1 U_2^* \rangle = q_1q_2\int_A\int_A \exp \left[ -{1\over2}\sigma^2(q_1^2-2\rho(x-x')q_1q_2+q_2^2)\right ] dx dx' .
\eeq
In the following we  use a simple approximation, 
\beq
\rho(u)=1-|u|/l, \: \: |u| \leq l,
\eeq
 else zero. In this simple  case  we model  a Gaussian correlation function by a linear approximation. For simplicity we will also work in the 1-D case. 
The first step now is to define
$
u=x-x'
$, and $v=x+x'$. The Jacobian of this change of variables is $1/2$. The integral becomes (let $A$ be a 1-D area from $-L$ to $L$) 
\bea
\langle U_1 U_2^* \rangle &=& q_1q_2 \int_{-\sqrt{2}L}^{\sqrt{2}L} \int_{-\sqrt{2}L+u}^{\sqrt{2}L-u} {du\,dv\over 2}\,  \exp \left[ -{1\over2}\sigma^2(q_1^2-2\rho(u)q_1q_2+q_2^2)\right ]\\
&=& \sqrt{2}L q_1q_2 \int_{-\sqrt{2}L}^{\sqrt{2}L} du\, \exp \left[ -{1\over2}\sigma^2(q_1^2+2\rho(u)q_1q_2+q_2^2)\right ].
\eea
Now, using our simplified correlation function model (and we will henceforth assume that $L>l$), we find
\beq
\langle U_1 U_2^* \rangle= 2\sqrt{2}L \left( {l\over \sigma^2 } \left( 1-e^{-\sigma^2 q_1q_2}\right) e^{-\sigma^2\Delta q^2 /2}  +q_1q_2(\sqrt{2}L-l) e^{-\sigma^2(q^2_1+q_2^2)/2} \right).
\eeq
For L-band, $q=2k$ is about 60 per meter. If $\sigma$ is greater than 0.1 m, $\sigma^2*q^2$ is greater than 40, and all the exponential terms in this expression vanish, except for $\exp[{-\sigma^2\Delta q^2 /2}]$. That is, for $\sigma >  0.1 $ m, 
 \beq
\langle U_1 U_2^* \rangle\sim {2\sqrt{2}L l\over \sigma^2 }  e^{-\sigma^2\Delta q^2 /2} .
\eeq
It also follows that 
\beq
C(U_1,U_2) = { \langle U_1 U_2^* \rangle \over \sqrt{ \langle U_1 U_1^* \rangle \langle U_2 U_2^* \rangle}} \sim  e^{-\sigma^2\Delta q^2 /2} .
\eeq
According to this result, the critical parameter in any reasonable ocean state is simply the significant wave height. Imagine for instance an ocean with a very large correlation length and large height standard deviation. This is mirror-like ocean with a global up-down displacement. It is quite clear that ih this example $\langle U_1 U_2^* \rangle$ will be zero, since this product will be a number in the complex plane with a reasonable magnitude and a rather random phase.  The average of such a complex number is zero. 

The general result, relevant for smaller significant wave heights, is
\beq
C(U_1,U_2) = {{l\over \sigma^2 } \left( 1-e^{-\sigma^2 q_1q_2}\right) e^{-\sigma^2\Delta q^2 /2}  +q_1q_2(\sqrt{2}L-l) e^{-\sigma^2(q^2_1+q_2^2)/2}
\over
\sqrt{ \left(
{l\over \sigma^2 } \left( 1-e^{-\sigma^2 q_1^2}\right)   +q_1^2(\sqrt{2}L-l) e^{-\sigma^2q^2_1} \right)
 \left(
{l\over \sigma^2 } \left( 1-e^{-\sigma^2 q_2^2}\right)   +q_2^2(\sqrt{2}L-l) e^{-\sigma^2q^2_2} \right)
}}.
\eeq 
 We show a plot for the case $l=5$ and $L=20$ in Figure \ref{coherency}.

 Although we did not discuss it, it is rather immediate that the average field for the pure fields are governed again by $\exp{[-\sigma^2q^2 /2]}$---this can be read off Equation~\ref{average} assuming a zero correlation function to infer $\langle \exp {[iq z(x)]} \rangle = \exp{[-\sigma^2q^2 /2]}$.

We have checked this trend in our data. In general, the interferometric combination average fields are larger.   For instance, for a  $\sigma_\zeta$= 79 cm simulation (50 seconds at 0.0025 temporal resolution, 44 m size, spatial resolution 10cm, U10=1100cmpers)---see Figure~\ref{comparison1100}, the  L1, L2, L5, L12, L25 average fields were: 
%\begin{itemize}
%\item[L1:] ( 0.035127546,   -0.0081386915) \\[-0.8cm]
%\item[L2:] ( 0.0011217963,    -0.061512818) \\[-0.8cm]
%\item[L5:](    -0.036160155,     0.026374968) \\[-0.8cm]
%\item[L12:](    -0.014956629,     0.087532800)\\[-0.8cm]
%\item[L25:](     -0.12120218,     0.070742104)  \\[-0.8cm]
%\end{itemize}
%while for U10=4 ms, 
%\begin{itemize}
%\item[L1:](     0.052441531,    -0.033514758)\\
%\item[L2:](    -0.064432181,     -0.11885066) \\
%\item[L5:](    -0.035560077,      0.10472799) \\
%\item[L12:](    -0.090138286,      0.11579234) \\
%\item[L25:](     -0.630366530,      0.20110769)\\
%\end{itemize}
\begin{quote}
\begin{itemize}
\item[L1:] $ ( +0.03,   -0.01) \\[-0.6cm] $
\item[L2:] $ ( +0.00,    -0.06) \\[-0.6cm] $
\item[L5:] $(    -0.04,     +0.03) \\[-0.6cm] $
\item[L12:] $(    -0.02,     +0.09)\\[-0.6cm] $
\item[L25:] $(     -0.12,     +0.07)  \\[-0.3cm] $
\end{itemize}\end{quote} 
while for U10=4 ms,  $\sigma_\zeta$= 12 cm simulation, 19 m side (see Figures~\ref{coherence400.ps} ,  \ref{194pix2,r=10cm,U10=400fieldpoints.ps} and  \ref{histo.ps}  for more details about  this simulation) \begin{quote}
\begin{itemize}
\item[L1:] $(     +0.05,    -0.03)\\[-0.6cm] $
\item[L2:] $(    -0.06,     -0.12) \\[-0.6cm] $
\item[L5:] $(    -0.04,      +0.10) \\[-0.6cm] $
\item[L12:] $(    -0.09,      +0.12) \\[-0.6cm] $
\item[L25:] $(     -0.63,      +0.20)\\[-0.3cm] $
\end{itemize}
\end{quote}
This is where the real strength of PIP becomes evident. Moreover, low-pass filtering (see next section) does not change these numbers much---the average field is robust.

The Fraunhofer approximation has given us a useful means to understand the correlation properties of the signals in different frequencies. We should note, however, that in the GPS case we are not in the Fraunhofer zone---we are in the WAF zone, which is is usually substantially larger.  Understanding all the implications of the extension to the WAF zone involves the analysis of the ``beast'' in Equation \ref{beast} which is substantially more complicated by the factors inthe areal function. This is left for future work.

\section{Filtering}
\begin{figure}[b!] 
\epsfxsize=150mm 
\epsffile{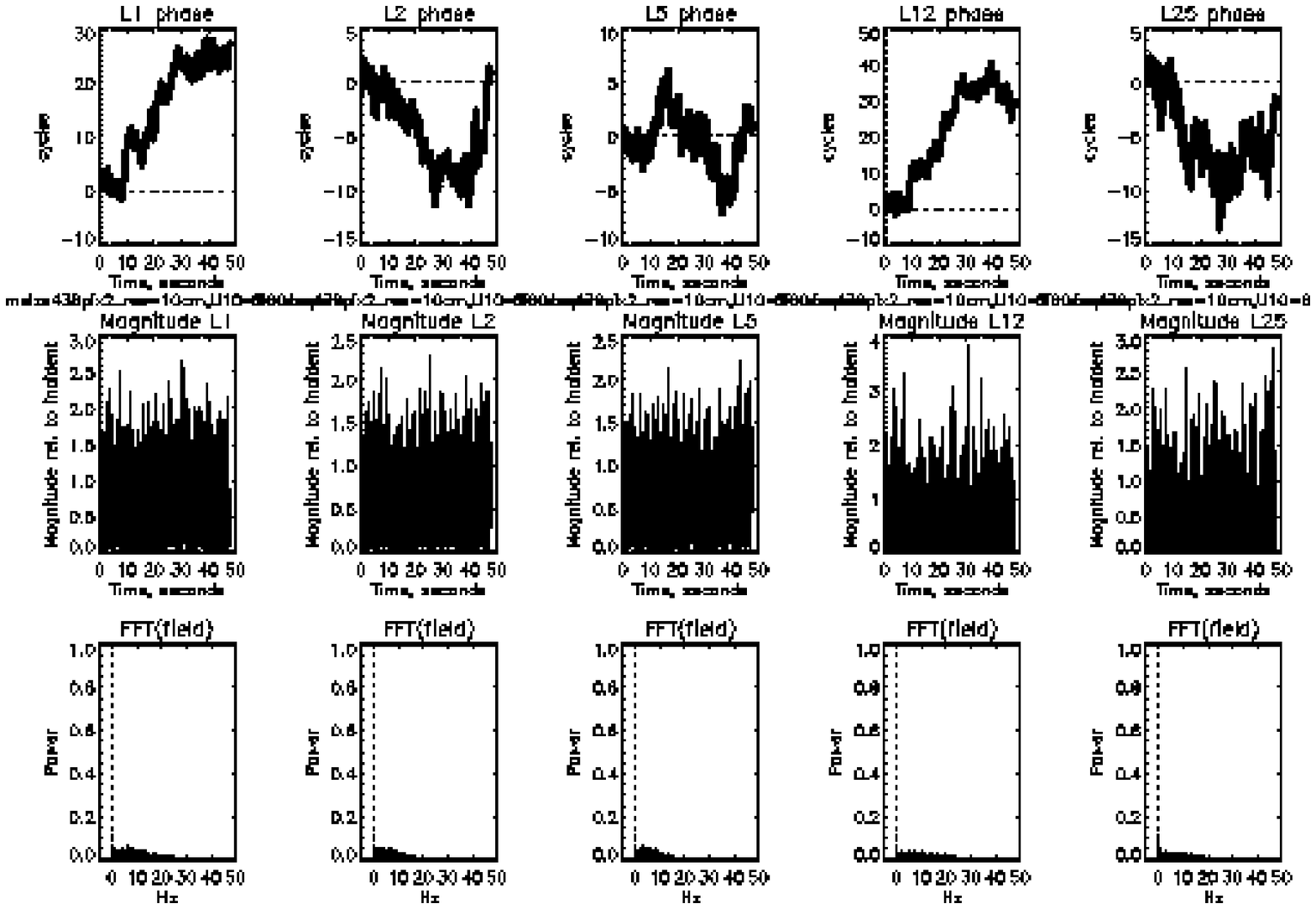}
\epsfxsize=150mm 
\epsffile{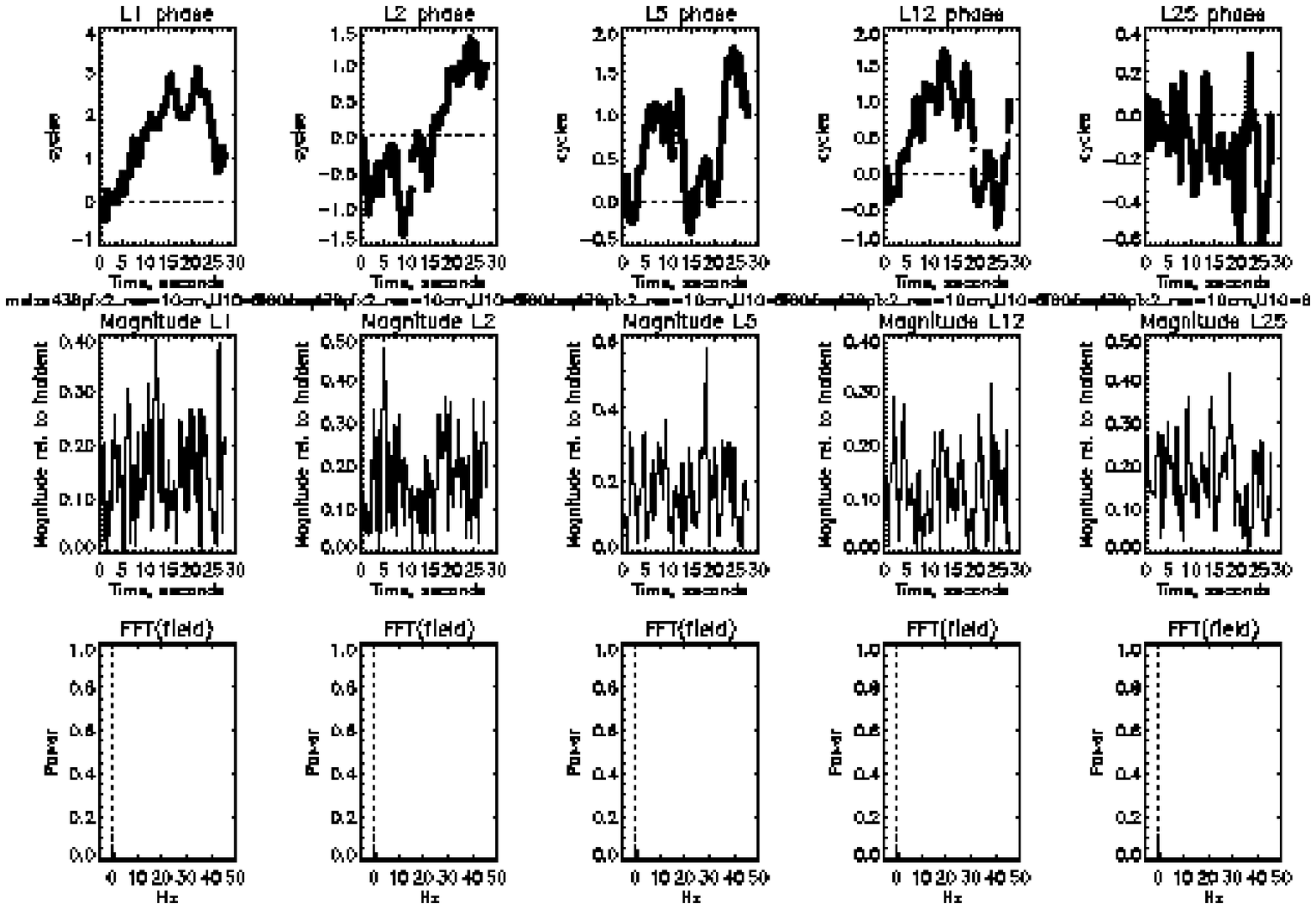}
\caption{ \label{comparison} These are plots of the fields and phases for the case U10=6 m/s  ($\sigma_\zeta=42$ cm), with an ocean patch of 43 m, resolution 10 cm. The top is the original time series, the second is with a field filter at 0.5 Hz.   } 
\end{figure}

Another way to understand and take advantadge of the coherence proporties is to focus on the low frequency aspects of the reflected signals for the static case. As we mentioned above, in the static situation the coherence function of the scattered phase is related to the zeroth Fourier component of the field---see Equation \ref{cohfcn}. Filtering the signal to retrieve slow varying geophysical signals is thus a plausible approach. As an example, we show in Figure \ref{comparison} the fields with and without filtering at 0.5 Hz (relevant for the case of a LEO). The result is quite good, as can be seen. This behavior is also expected from the analysis above of the average field above, of course.  The interferometric combination is again superior, especially L$_{25}$. For the satellite of aircraft case, good modeling of the receiver position and filtering  can be used to remove all but the slowly varying geophysical signals from the phase drift. 

\begin{figure}[b!] 
\epsfxsize=150mm 
\epsffile{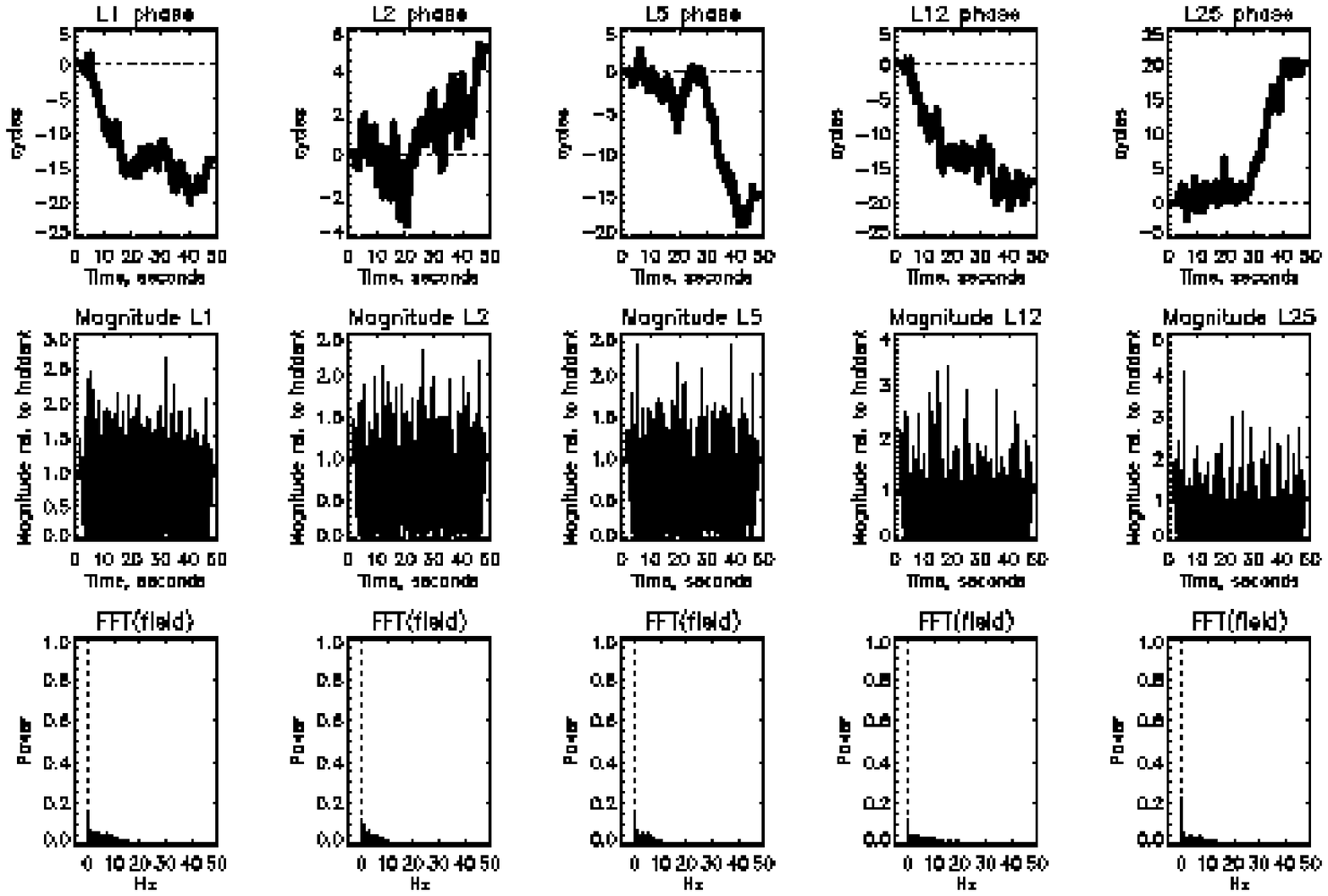}
\epsfxsize=150mm 
\epsffile{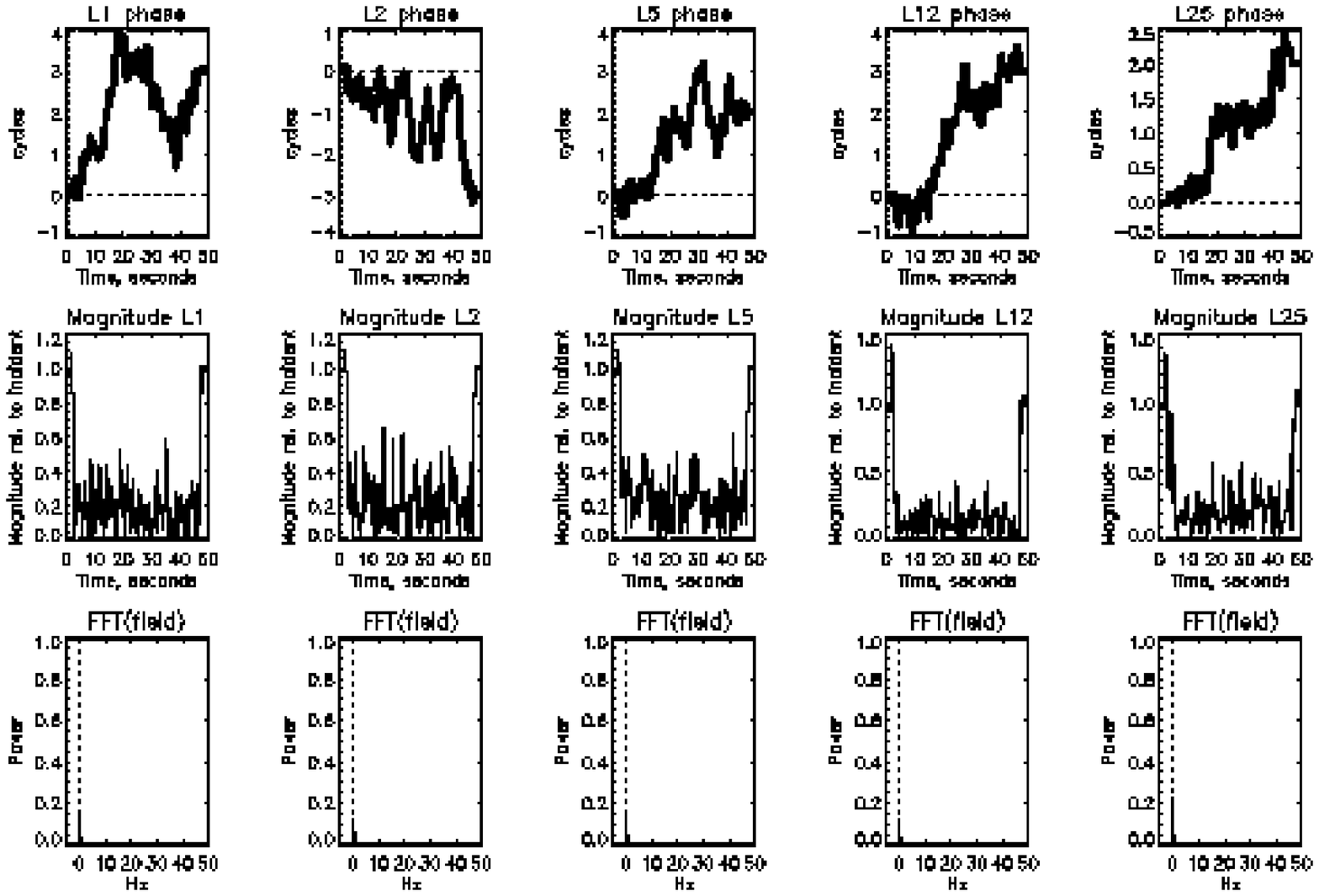}
\caption{ \label{comparison300} These are plots of the fields and phases for the case U10=3m/s ($\sigma_\zeta=18$ cm), with an ocean patch of 43 m, resolution 10 cm. The top is the original time series, the second is with a field filter at 0.5 Hz.   Temporal resolution in all the simulations  is of 0.0025 seconds. } 
\end{figure}

\begin{figure}[b!] 
\epsfxsize=150mm 
\epsffile{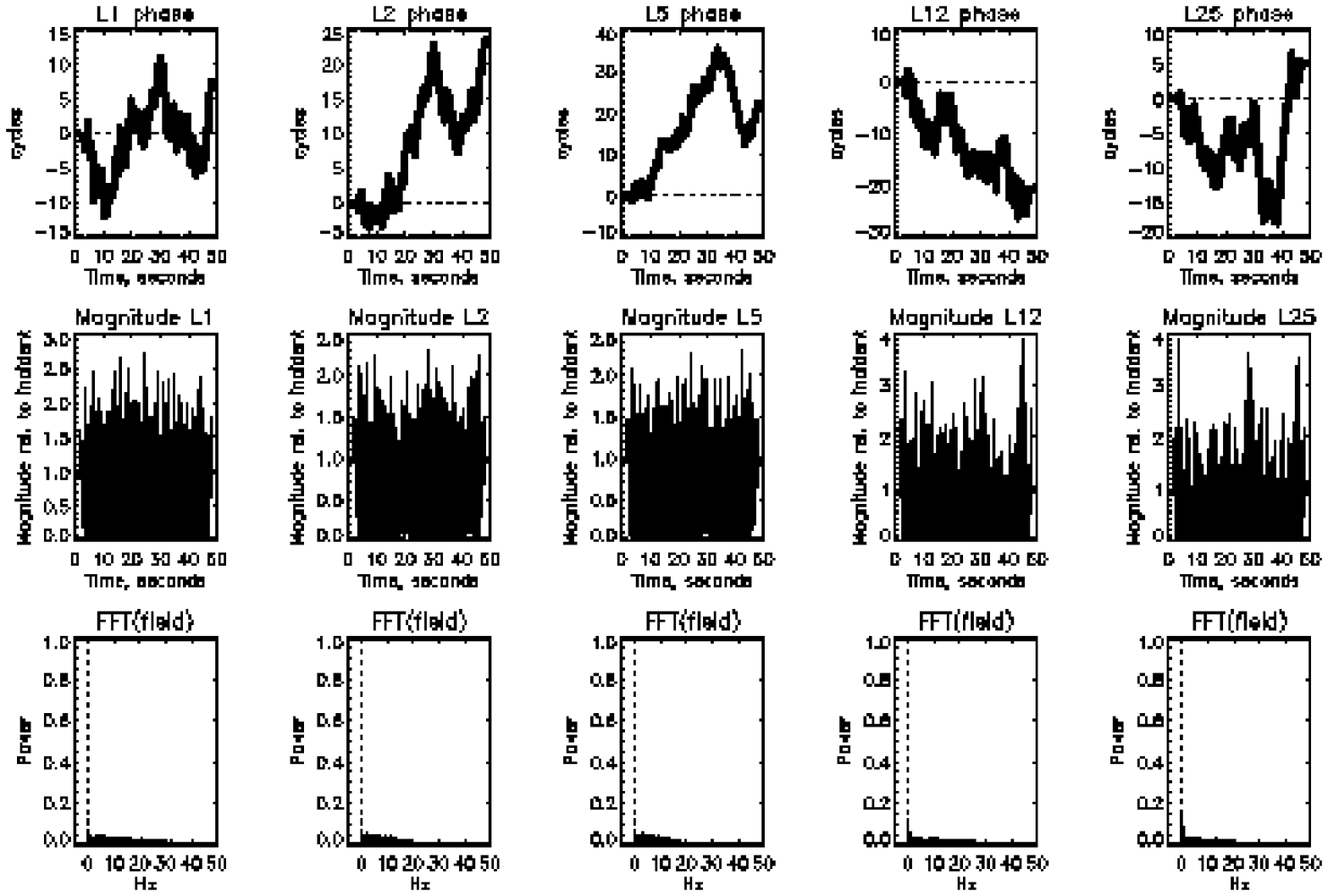}
\epsfxsize=150mm 
\epsffile{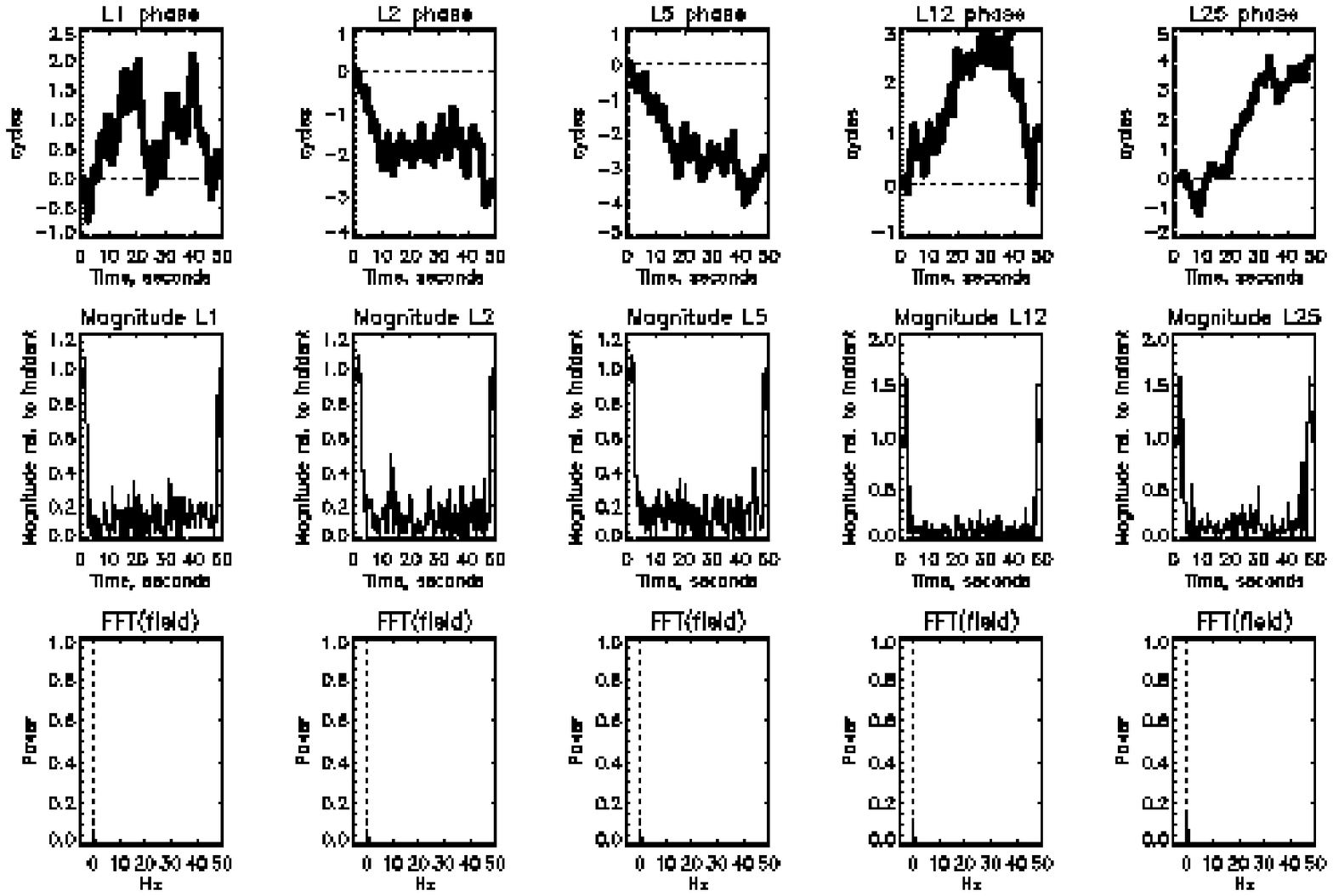}
\caption{ \label{comparison1100} These are plots of the fields and phases for the case U10=11m/s ($\sigma_\zeta=79$ cm), with an ocean patch of 43 m, resolution 10 cm. The top is the original time series, the second is with a field filter at 0.5 Hz. See the next figure for the corresponding (unfiltered) coherence and strucuture functions. This is complete simulation, with the ocean going from flat to moving to flat. Observe the picked up winding number.  Filtering at 0.5 Hz did not entirely cure the problem in this case. } 
\end{figure}

Important points: filtering doesn not seem to affect the field mean value (a very desirable result). It does significantly reduce the scatter, however. For instance, in Figure \ref{comparison300}, we have the fields for a 3 m/s wind. The standard deviations of the field go from 0.87, 0.82, 0.79, 0.73, 0.68 to 0.51, 0.51, 0.49, 0.48, 0.29 (for L1, L2, L5, L12, L25) after 0.5 Hz filtering of the fields. See also the results for higher wind speed is Figures \ref{comparison} and \ref{comparison1100}.  The mean field remains virtually unchanged after this process.

%&\approx& q_1q_2       {1\over i\Delta q} 2 \sin L \Delta q  \int \langle {\cal A}_1'(r) {\cal A}_2'(r+ \varepsilon) \rangle_t \, e^{i(q_1+q_2)\varepsilon/2}\,  d \varepsilon
%\eean
%where in the last step we have assumed that the areal function correlation function depends only on the separation bewtween measurement points.  Thus, we see that for the fields to be correlated we need a small contributing area.  If we want to use P-code ships, the we need a 30 meter synthetic wavelenght. For C/A code, the frequencies need to be 1 MHz apart. There is another term in the above equation: the Fourier transform of the areal function correlation fucntion also appears.

\chapter{Simulation  experiments with winding number and GIP}
Consider the following scenario. The receiver is static.  Initially the ocean is calm, totally flat. Then a disturbance gradually appears, peaks, and slowly disappears. The ocean final state is the same as the original state: total flatness. We have run several simulations with this scenario. The result is that the winding number of the field history is non-zero. That is, the field can loop several times arond zero.  

\begin{figure}[h!] 
%\hspace{2.5cm} 
\epsfxsize=150mm 
\epsffile{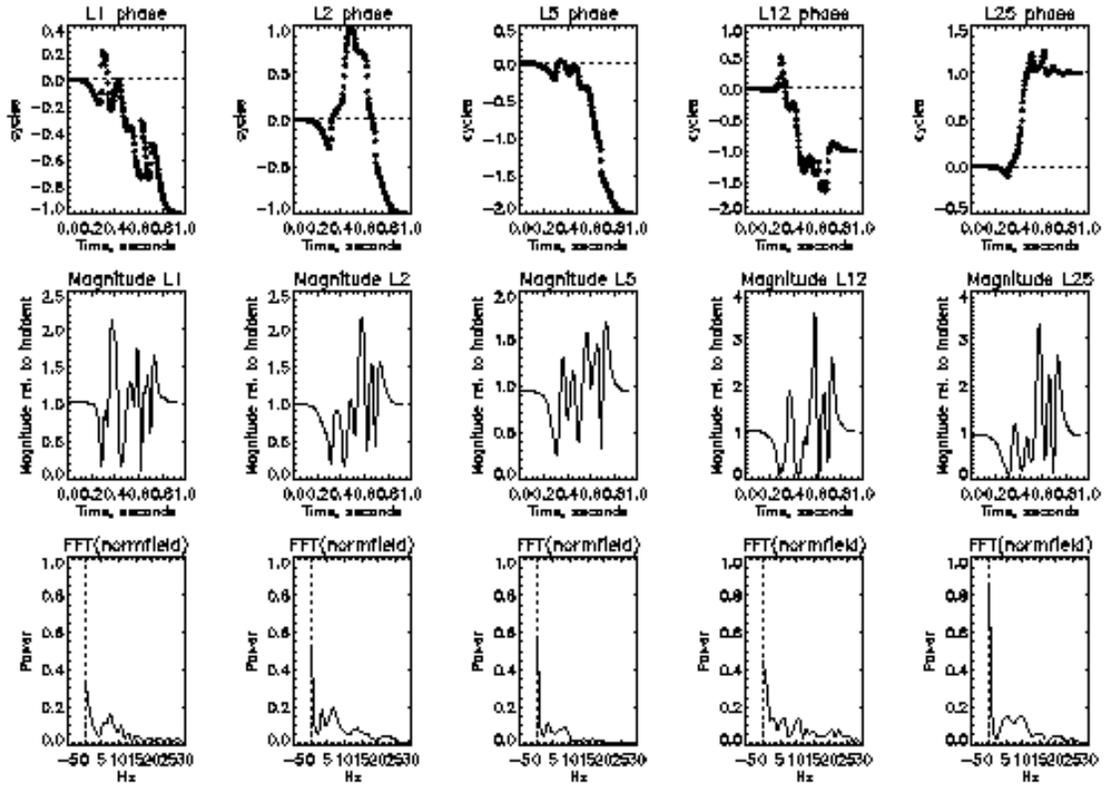} 
\caption{ \label{special1} Simulation: 20 meters side, res=10cm, max ocean 4 m/s U10 with a peak $\sigma_\zeta$ of 12 cm, sine fourth modulation. Note that in the bottom row we present the FFT  power of the normalized field. The zero component is basically the coherence function mentioned earlier. The interferometric component shows a higher degree of coherence. } 
\end{figure}

\begin{figure}[h!] 
\hspace{2.5cm} 
\epsfxsize=90mm 
\epsffile{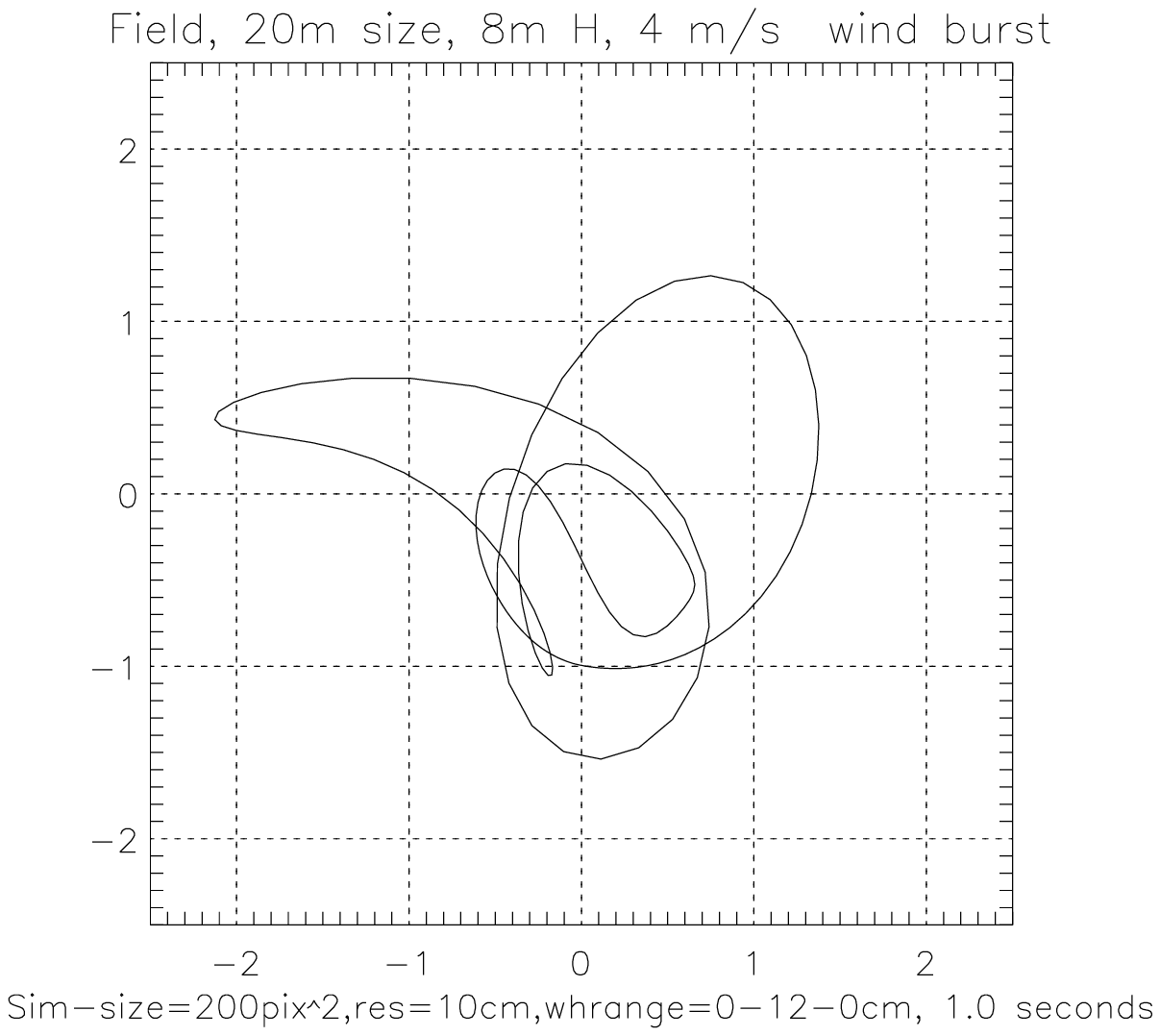} 
\caption{ \label{special1w} In this Figure we illustrate  the L2 field phasor in Figure~\ref{special1}, which has a winding number   of -1.  } 
\end{figure}

\begin{figure}[h!] 
\hspace{2.5cm} 
\epsfxsize=90mm 
\epsffile{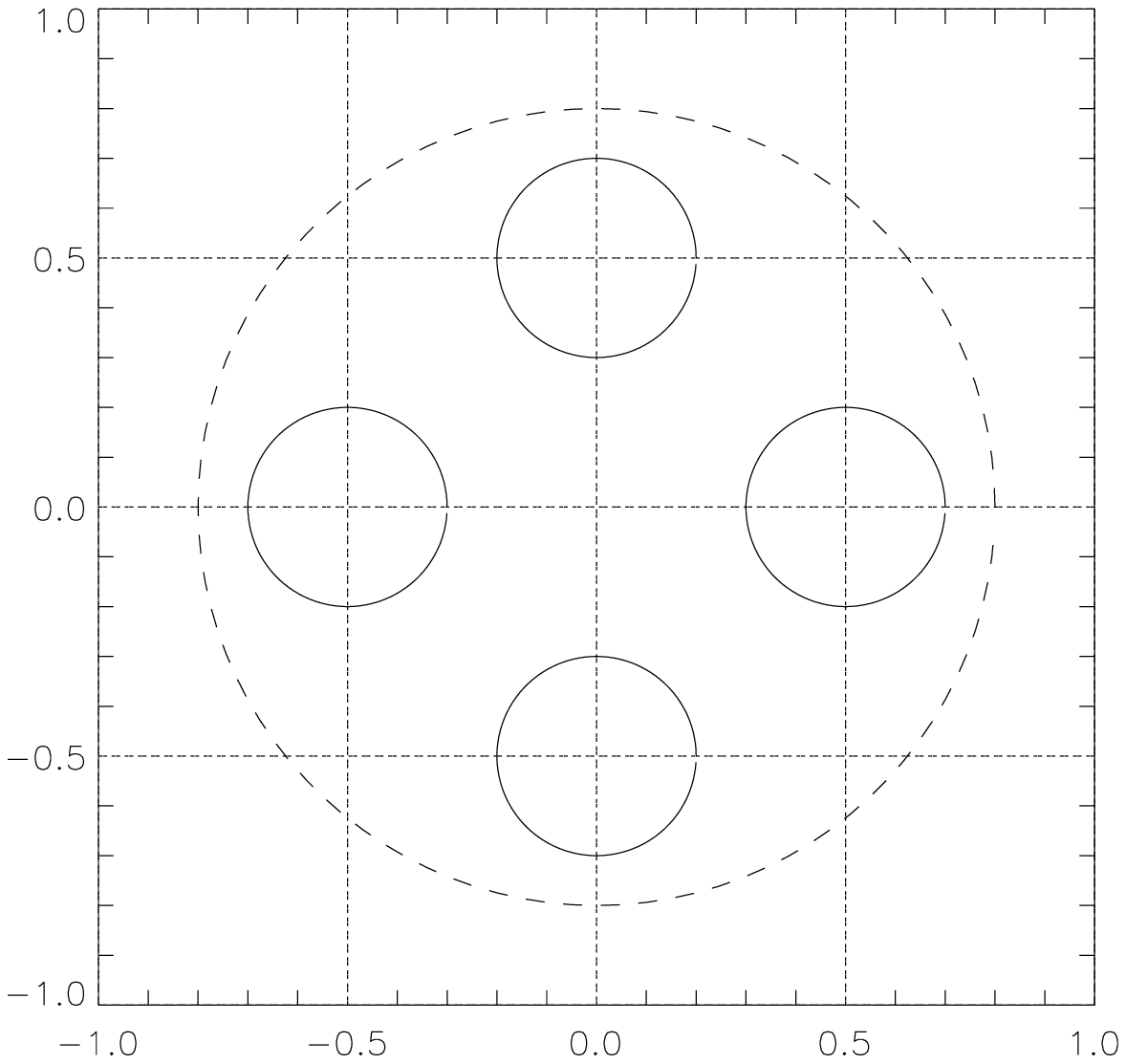} 
\caption{ \label{circles} Winding number magic: four curves with winding number 0 yield a winding number 1 curve after being added. The trick can be done with 2 curves. In this case they all turn in the same direction. Changing that will result in an ellipse.} 
\end{figure}

The first example is in Figure \ref{special1} and accompanying  Figure \ref{special1w}.
How is this possible?  As can be seen in the example in  Figure \ref{circles}, the sum of zero-winding-number curves can result in a  winding number 1 curve. Since the resulting total  field is a linear combination of such fields, all that we need to show is how the moving ocean can generate such winding number zero curves.  This is easy: the only requirements are that a) the initial and final phase be the same, with the phase moving in between, and b) that the field magnitude increase or decrease before returning to its original value. It is possible to imagine  examples based on GO reasoning (a usefull approximation, at any rate) to obtain this behaviour. Suppose, for instance, that the distance from a specular point to the receiver decreases and then increases, while the radius of curvature decreases then increases. It is easy to see that this yields a clockwise loop in phasor space (with the usual angular convention used in the complex plane).  
   
%%%%%%%%%%%%%%%%%%%%%%%%%%%%%%%%%%%%%5555
%\begin{figure}[h!] 
%%\hspace{2.5cm} 
%\epsfxsize=150mm 
%\epsffile{Sim-size=438pix^2,res=10cm,whrange=10cm.PIPhase.ps} 
%\caption{ \label{special2} Simulation: 44 meters side, res=10cm, an ocean of 2  m/s u10. This is about   0.8 meters peak to trough.} 
%\end{figure} 

%\begin{figure}[h!] 
%%\hspace{2.5cm} 
%\epsfxsize=150mm 
%\epsffile{Sim-size=438pix^2,res=10cm,whrange=72cm.PIPhase.ps} 
%\caption{ \label{special10} Simulation: 44 meters side, res=10cm, an ocean of 10  m/s u10. This is more than 4 meters peak to trough.} 
%\end{figure} 

%\begin{figure}[h!] 
%%\hspace{2.5cm} 
%\epsfxsize=150mm 
%\epsffile{Simsize194pix2_res=10cm,U10=1500cmpersPIPhase.ps} 
%\caption{ \label{special15} Simulation: 20  meters side, res=10cm, an ocean of 15  m/s u10. This is more than 6 meters peak to trough in the real situation. Due to a bug in the ocean generating code (it is able to include only long waves in the power if the simulation size is large), we have a situation here of 50 cm std. } 
%\end{figure} 

%\begin{figure}[h!] 
%%\hspace{2.5cm} 
%\epsfxsize=150mm 2
%\epsffile{Simsize194pix2_res=10cm,U10=500cmpersPIPhase.ps} 
%\caption{ \label{special5} Simulation: 20  meters side, res=10cm, an ocean of 5  m/s u10. Note that in the bottom row we show the FFT of the field, not of the normalized field ad in the previous ones. Note also the large zero component in the field---a sign of coherence.} 
%\end{figure} 
%%%%%%%%%%%%%%%%%%%%%%%%%%%%%%%%%%%%%%%%%%%%%%%%%%%%%%%%55
We have seen the phase dispersion effect, very much related to winding number (which is a special case of the first when initial and final states are the same), in all wind conditions, with U10=2 m/s  and up. At first sight this poses a significant problem in the use of phase for altimetry. How can we use a ``defiting'' measure for altimetry? It is now very important to characterize the statistical and geophysical properties of this phase drift (Geophysically Induced Phase drift, or GIP for short).  

\section{The uses of winding number}
Winding number, or more generally, phase dispersion under a moving ocean is certainly related to sea state characteristics. 
An interesting Gedanken\footnote{Gedanken (thought) experiments are little mental exercises for exploring hypothetical experiments. The terminology comes from  Einstein's work, who  is
   said to have engaged in them even when he was very young.}: imagine a satellite tracking a reflection over a smooth region, then rough, then smooth again. The winding number induced will be related to the roughness encountered.  We have already seen that, in general, the phase drifts without a specific chirality in our simulations. It remains to be seen if non-gaussian effects in the surface can lead to chirality, or preferred rotation direction. If this is the case, this phenomenon could conceivably be used for geophysical measurements.
We have already seen that GIP  is related to wind speed in our simulations.

\chapter{Recommendations for Post-processing Procedures}

%Based on the Elfouhaily et al model \cite{Elfouhaily97}, we have verified that ocean vertical speeds can be substantial (e.g, $\pm$10 m/s for a 20 m/s ocean).
%This implies Doppler shifts as high as 100 Hz on some components of the resulting field. At first sight, this means that in such situations coherent integration time cannot exceed 10 ms. Note that this reasoning is independent of the height of the receiver. It does assume a static receiver. The case of a moving receiver needs to be studied carefully, as the receiver motion induces other Doppler effects.  We have seen, however, that the coherence time for the interferometric signal is rather long---in fact, using the standard definition for coherence \cite{thomson}, the interferometric signal remains coherent for a time  long in a variety of situation. 

%Similarly, the Nyquist sampling rate for (correlator output) the received signal woud have to be around 200 Hz, whixh allows for 5 ms integration times. 
Based on the work carried out so far, we are prepared to outline some of the possible incarnations of the PIP concept. What we must emphasize is that the strength lies in the noise cancellation in the interferometric combination. Filtering is best carried out after this combination.

\section{The PIP instrument: architecture and data products}

Based on our results, we have the following suggestions.
\begin{enumerate}
\item Use two wideband (200 Hz for the static receiver, possibly more for the moving case) PLL's to extract the phase in each frequency. 
\item Combine the phases and  low-pass filter to recover altimetric trends. The normalized interferometric field is more coherent, so it can be used to track slow-varying altimetric changes (such as a tide).  We have seen that the interferometric fiels is a more robust source, with a better defined phase.
\item Use phase drift as a geophysical parameter  More generally, the phase and power time series are probably rich information source of sea state. 
\end{enumerate}
To be more precise, the following recipes for PIP implementation can be conceived: 
One, the most natural one perhaps, is the Single PLL PIP (SPIP)
\begin{enumerate}
\item   Multiply the two signals, form the interferometric signal.
\item   Filter from 0 to 0.5 Hz
\item   Use a single PLL to extract the phase.
\end{enumerate}
The disadvantadge is that the field magnitude will oscillate quite a bit. This may not be important, given that a filter will be used. 
An alternative would be to used two PLLs (MPIP)
\begin{enumerate}
\item   Extract the phase in L1 and in L2
\item   Combine the two phases
\item   Filter from 0 to 0.5 Hz.
\end{enumerate}
This is probably similar to the previous case, but it takes the field magnitude out of the picture from the very beginning---probably not a good thing. Finally, a non-contender is to  
\begin{enumerate}
\item   Filter from 0 to 0.5 Hz
\item   Extract the phase in L1 and in L2
\item   Combine the two phases
\end{enumerate}
This is a bad option, it makes no use of the jitter cancellation across frequencies.

We have focused on the static case. For other situations we just need to work with excess phase and carry out the same filtering process. 

%Finally, we should mention that it may be useful to make use of all available frequencies or more clever ways of combining them.

\chapter{Conclusions and Future Work}
\section{What we have found: PIP's superior performance}
\begin{enumerate}
\item The mean field is largest for the PIP  case L$_{25}$ for all sea conditions. In general the mean field is proportional to $\exp{[-\sigma^2q^2 /2]}$, where $q=4\pi/\lambda$ and $\lambda$ is the pure or interferometric wavelength.
\item The coherence integration time is consequently substantially larger for the interferometric combination, especially L$_{25}$ (if the mean sea level scatter is very large the effect disappears, of course). The interferometric phase is  coherent in a reasonable range of sea conditions, unlike the single frequency phases.
% We have seen this by plotting the resulting interferometric field in the complex plane: while for the single frequency case the plots are, to the eye, basically uniformly distributed around zero, for the interferometric field this is not so. The field has a  mean (i.e., both the modulus and the phase have a well-defined mean). Although for the single-frequency fields the coherence function does not vanish, the interferometric degree of coherence is clearly higher.  
This  implies that the PIP interferometric combination is superior  to extract altimetric low frequency trends in the phase, as we discussed at the beginning. 
%\item The correlation between the fields at different frequencies is much larger for L$_{25}$ than for L$_{12}$.  This is another way to see the coherence that results in the interferometric case.
\item The phase in our simulations behaves like a random walk, and the drift rate is directly related to sea state: in general, the larger the wind speed, the higher the drift rate. This implies that a system capable of tracking the phase of the reflected signals can provide, aside from altimetric measurements, sea state information.
\item Filtering the received field does not affect its mean value, while decreasing its scatter. It can probably be used for altimetric purposes, as it removes noise but leaves the slow-varying geophysical signals alone.
\item No long-term trends in phase drift chirality have been detected in our simulations with  Gaussian ocean models based on  \cite{Elfouhaily97}. We have performed simulations up to 2 minutes long. This is good news for our altimetric efforts.  It seems that despite the phase drift present, and the winding number it can lead to, in practice this effect should not introduce systematic effects on altimetric determination. It is important to check that the simulations are correctly imitated by nature, of course. Among other things, non-Gaussian effects in the sea could conceivably lead to real drifting.
   \end{enumerate}
 The single  most important parameter in the correlation between the fields at different frequencies is the significant wave height. This has to be compared to the synthetic wavelength. For rough seas, the correlation bethween the fields (and therefore interferometric coherence) in different frequencies disappears, and the coherence time goes to zero even for the interferometric combinations.  In calmer ocean conditions, however, our results indicate that the interferometric combination remains coherent. 
 \section{What we'd now like to know: future work}
Due to the limited scope of this study we were not able to cover all the interesting aspects of this problem. 
Future theoretical work includes the extension of the present results to the case in which the receiver is far away from the ground and is moving. We would also like to understand the impact of changing the ocean statistics from gaussian to non-gaussian. 
Another important aspect that deserves further research is to extend the analysis to  a bistatic situation. So far, our simulations have only considered the monostatic case.
We intend to continue this work within the scope of PARIS-$\alpha$:
\begin{enumerate}
\item Examine more realistic situations (higher, faster).  This includes, in particular, understanding the issues that will arise in the aircraft and LEO scenario, and it will require larger simulations---in fact it may  impossible to simulate the ocean with the required size and resolution, as the WAF zone are increases linearly with height. 
\item Understand the theoretical issues involving the  correlations and Doppler spread of the reflected fields without using the Fruanhofer approximation for the receiver.
\item Extend the analysis to the bistatic situation. 
\item Carry out a detailed study of phase drift versus wind speed. 
\item   Look deeper at the advantages (which we have seen and are expected) of using even closer frequencies. This may result in recomendations for GALILEO, for instance.
%\item We would like to produce a table which, given set of wind speeds, produces the mean field, mean field scatter, drift rate, coherence time, the correlation between fields, and the revised values of all these after filtering.   We have developed the ``technology'' to do this work, but time has been too short.  We believe, however, that the key ideas are in this report.
\item Understand the impact of non-Gaussian ocean effects.
\item Understand better the possible sources of chirality in GID. This is important to assure altimetric accuracy.
\item Obtain and analyze data for the static situation. 
\item Understand the effectiveness of GO or its higher order corrections.   
\end{enumerate}
It is rather clear that experimental work is needed in this field. The most important experiment for the concept at this point, is a redo of the bridge experiment: a static receiver over the moving ocean, using both L$_1$ and L$_2$. The desired product from such an experiment from the point of view of the present work, would  time-series' of the fields.  
MEATEX campaigns will also provide aircraft data which will be very useful to understand some of these issues. 

\chapter{Acknowlegments} 
This is a document produced for ESA under ESTEC Contract No. 14071/99/NL/MM. The author's are grateful to ESA for funding and permission to freely distribute this document, and are especially grateful to Manuel Martin-Neira of ESTEC, the technical officer in charge of this contract, for very valuable comments and suggestions during the course of this research. 
. 
   %}
 %%%%%%%%%%%%%%%%%%%%%%%%%%%%%%%%%%%%%%%%%%%%%%%%%%%%%%%%%%%%%%%%%%%%%
\end{document}